\documentclass[11pt]{article}
\usepackage{amssymb,amsmath,amsfonts}
\usepackage{graphicx}
\usepackage{graphics}
\usepackage{eepic,epsfig}

\@addtoreset{equation}{section}

\makeatother

\begin{document}

\title{Induced fermionic charge and current densities\\
in two-dimensional rings}
\author{S. Bellucci$^{1}$\thanks{%
E-mail: bellucci@lnf.infn.it },\, A. A. Saharian$^{2,3}$ \thanks{%
E-mail: saharian@ysu.am},\, A. Kh. Grigoryan$^{3}$ \thanks{%
E-mail: ashot.gr@gmail.com} \\
\\
\textit{$^1$ INFN, Laboratori Nazionali di Frascati,}\\
\textit{Via Enrico Fermi 40, 00044 Frascati, Italy} \vspace{0.3cm}\\
\textit{$^2$Department of Physics, Yerevan State University,}\\
\textit{1 Alex Manoogian Street, 0025 Yerevan, Armenia}\vspace{0.3cm}\\
\textit{$^3$Institute of Applied Problems in Physics NAS RA,}\\
\textit{25 Nersessian Street, 0014 Yerevan, Armenia}}
\maketitle

\begin{abstract}
For a massive quantum fermionic field, we investigate the vacuum expectation
values (VEVs) of the charge and current densities induced by an external
magnetic flux in a two-dimensional circular ring. Both the irreducible
representations of the Clifford algebra are considered. On the ring edges
the bag (infinite mass) boundary conditions are imposed for the field
operator. This leads to the Casimir type effect on the vacuum
characteristics. The radial current vanishes. The charge and the azimuthal
current are decomposed into the boundary-free and boundary-induced
contributions. Both these contributions are odd periodic functions of the
magnetic flux with the period equal to the flux quantum. An important
feature that distinguishes the VEVs of the charge and current densities from
the VEV of the energy density, is their finiteness on the ring edges. The
current density is equal to the charge density for the outer edge and has
the opposite sign on the inner edge. The VEVs are peaked near the inner edge
and, as functions of the field mass, exhibit quite different features for
two inequivalent representations of the Clifford algebra. We show that,
unlike the VEVs in the boundary-free geometry, the vacuum charge and the
current in the ring are continuous functions of the magnetic flux and vanish
for half-odd integer values of the flux in units of the flux quantum.
Combining the results for two irreducible representations, we also
investigate the induced charge and current in parity and time-reversal
symmetric models. The corresponding results are applied to graphene rings
with the electronic subsystem described in terms of the effective Dirac
theory with the energy gap. If the energy gaps for two valleys of the
graphene hexagonal lattice are the same, the charge densities corresponding
to the separate valleys cancel each other, whereas the azimuthal current is
doubled.
\end{abstract}

\bigskip


\bigskip

\section{Introduction}

Field theories for the number of spatial dimensions $D$ other than 3 have
attracted a great deal of attention. For $D>3$ this was mainly motivated by
the importance of the Kaluza-Klein and braneworld type models, as frameworks
for the unification of the fundamental physical interactions. Extra
dimensions are an inherent feature in string theories and in supergravity.
There has been a growing interest in recent years in models formulated on
backgrounds with the number of spatial dimensions $D<3$. Aside from their
role as simplified models in particle physics, field theories in lower
dimensions serve as effective theories describing the long-wavelength
properties of a number of condensed matter systems \cite{Frad91,Gusy07}.
Examples for the latter are high-temperature superconductors, d-density-wave
states, Weyl semimetals, graphene (and graphene related materials) and
topological insulators. For these systems, the long-wavelength dynamics of
excitations is formulated in terms of the Dirac-like theory living in
(2+1)-dimensional spacetime where the role of the velocity of light is
played by the Fermi velocity. In topological insulators, $2D$ massless
fermionic excitations appear as edge states on the surface of a $3D$
topological insulator. (2+1)-dimensional models also appear as high
temperature limits of 4-dimensional field theories.

Among interesting features in (2+1)-dimensional models are flavour symmetry
breaking, parity violation, fractionalization of quantum numbers, the
possibility of the excitations with fractional statistics. Important new
possibilities appear in gauge theories. In particular, the topologically
gauge invariant terms in the action provide masses for the gauge fields.
This leads to a natural infrared cutoff in the theory and to the solution
for the infrared problem without changing the physics in the ultraviolet
range \cite{Dese82}. A possible mechanism for the generation of gauge
invariant topological mass terms is provided by quantum corrections \cite%
{Niem83}. The corresponding theories provide a natural framework for the
investigation of the quantum Hall effect. In models with fermions coupled to
the Chern-Simons gauge field, there are states with nonzero magnetic field
and with the energy lower than the lowest energy state in the absence of the
magnetic field \cite{Cea86}. As a consequence of this, the Lorentz
invariance is spontaneously broken \cite{Hoso93}. Among the most interesting
topics in the studies of (2+1)-dimensional theories is the parity and chiral
symmetry-breaking. In particular, it has been shown that a background
magnetic field can serve as a catalyst for the dynamical symmetry breaking
\cite{Klim92,Gusy94}. In addition, the background gauge fields give rise to
the polarization of the ground state for quantum fields with the generation
of various types of quantum numbers \cite{Niem83,Redl84}. In particular,
charge and current densities are induced \cite{Jaro86}-\cite{Li93}.

In a number of field theoretical models, including the ones describing the
condensed matter systems at large length scales, additional boundary
conditions are imposed on the field operator. These conditions can have
different physical origins. For example, in graphene nanotubes and
nanoloops, because of the compactification of one or two spatial dimensions,
the Dirac equation is supplemented by quasiperiodicity conditions along
compact dimensions with phases depending on the wrapping direction
(chirality of the nanotube). Another type of graphene made structures in
which additional boundary conditions are imposed on the field wavefunctions
are graphene nanoribbons, geometrically terminated single layers of graphite
(see, for instance, \cite{Been08}). The edge effects play a crucial role in
electronic properties of nanoribbons. In particular, depending on the
boundary conditions, a nonzero band gap may be generated. An important new
thing is the possibility for the appearance of dispersive edge states.

Among the most interesting physical consequences originating from the
spatial confinement of a quantum field is the Casimir effect \cite{Most97}.
The boundary conditions modify the spectrum of zero-point fluctuations and,
as a consequence of that, the vacuum expectation values (VEVs) of physical
observables are shifted. The physical quantities, most popular in the
investigations of the Casimir effect, are the vacuum energy and stresses. By
using these quantities, the forces acting on the constraining boundaries can
be evaluated. These forces are presently under active experimental
investigations \cite{Most97,Klim09}. For charged fields, among the most
important characteristics of the ground state are the expectation values of
the charge and current densities. Similarly to the vacuum energy and
stresses, the VEVs of these quantities are influenced by the change of
spatial topology or by the presence of boundaries. The vacuum currents in
spaces with nontrivial topology and with quasiperiodic boundary conditions
on the field operator along compact dimensions have been investigated in
\cite{Bell09} for the flat background geometry and in \cite{Bell13} and \cite%
{Beze15} for locally de Sitter and anti-de Sitter backgrounds. For the
special case $D=2$, the general results were applied to cylindrical and
toroidal graphene nanotubes, within the framework of the effective Dirac
theory. The influence of additional boundaries on the vacuum currents along
compact dimensions has been discussed in \cite{Bell15M} and \cite{Bell15AdS}
for locally Minkowski and anti-de Sitter backgrounds. The combined effects
of the topology, induced by a cosmic string, and coaxial boundaries on the
vacuum currents have been studied in \cite{Beze10,Beze15CS}.

In the present paper we investigate the VEVs of the fermionic charge and
current densities induced by a magnetic flux in a spatial region of
(2+1)-dimensional spacetime bounded by two concentric circles. On the
circles, bag boundary conditions are imposed. We assume that the flux is
located inside the inner boundary and, consequently, its effect on the
vacuum properties is of the Aharonov-Bohm type (for the influence of the
flux and boundaries on the vacuum energy see Refs. \cite{Lese98}). We have
organized the paper as follows. In the next section we specify the bulk and
boundary geometries and the boundary conditions imposed on the fermionic
field in the problem under consideration. In section \ref{sec:Modes}, the
complete set of positive- and negative-energy mode functions is determined
in the geometry of a finite width ring. These mode functions are used in
section \ref{sec:Charge} for the evaluation of the VEV\ of the charge
density. Two equivalent representations are provided with the explicitly
separated boundary contributions. The VEV\ of the azimuthal current density
is investigated in section \ref{sec:Current}. In section \ref{sec:Tsym},
based on the results from previous sections, the induced charge and current
densities are discussed in parity and time-reversal symmetric models and
applications are given for graphene rings. The main results of the paper are
summarized in section \ref{sec:Conc}. Throughout the paper the units with $%
\hbar =c=1$ are used, except for the part of section \ref{sec:Tsym} where we
discuss the applications to graphene rings.

\section{Problem setup}

\label{sec:Setup}

In a curved background geometry described by the metric tensor $g_{\mu \nu }$
and in the presence of an external electromagnetic field with the vector
potential $A_{\mu }$ the Dirac equation for a quantum fermion field $\psi (x)
$ is presented as
\begin{equation}
\left( i\gamma ^{\mu }D_{\mu }-sm\right) \psi (x)=0,  \label{Direq}
\end{equation}%
where $D_{\mu }=\partial _{\mu }+\Gamma _{\mu }+ieA_{\mu }$ is the gauge
extended covariant derivative, $\Gamma _{\mu }$ is the spin connection and $e
$ is the charge of the field quanta. The Dirac matrices $\gamma ^{\mu }$
obey the Clifford algebra $\{\gamma ^{\mu },\gamma ^{\nu }\}=2g^{\mu \nu }$.
As \ the background geometry, we consider (2+1)-dimensional flat spacetime
described in polar coordinates $x^{\mu }=(t,r,\phi )$ with the metric tensor
$g_{\mu \nu }=\mathrm{diag}(1,-1,-r^{2})$. It is well known that in odd
number of spacetime dimensions the Clifford algebra has two inequivalent
irreducible representations (with $2\times 2$ Dirac matrices in (2+1)
dimensions). Firstly we shall discuss the case of a fermionic field
realizing the irreducible representation of the Clifford algebra. The
parameter $s$ in Eq. (\ref{Direq}), with the values $s=+1$ and $s=-1$,
corresponds to two different representations (for more details see below).
With these representations, the mass term violates both the parity ($P$-)
and time-reversal ($T$-) invariances. The vacuum currents in the parity and
time-reversal symmetric models will be discussed in section \ref{sec:Tsym}.
In the long wavelength description of the graphene, $s$ labels two Dirac
cones corresponding to $\mathbf{K}_{+}$ and $\mathbf{K}_{-}$ valleys of the
hexagonal lattice.

\begin{figure}[tbph]
\begin{center}
\epsfig{figure=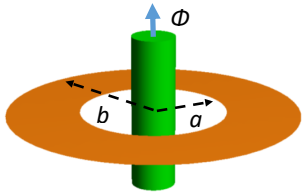,width=4.5cm,height=3cm}
\end{center}
\caption{Two-dimensional ring threaded by a magnetic flux.}
\label{fig1}
\end{figure}

We assume that the field is confined in the spatial region bounded by two
concentric circles having radii $a$ and $b$, $a<b$ (two-dimensional ring,
see figure \ref{fig1}). On the edges of this region the field operator obeys
the MIT bag boundary conditions
\begin{equation}
\left( 1+in_{\mu }\gamma ^{\mu }\right) \psi (x)=0,\;r=a,b.  \label{BCMIT}
\end{equation}%
with $n_{\mu }$ being the outward pointing unit vector normal to the
boundaries. In the region $a\leqslant r\leqslant b$ one has $n_{\mu
}=n_{u}\delta _{\mu }^{1}$ for the boundary at $r=u$ with%
\begin{equation}
n_{a}=-1,\;n_{b}=1.  \label{nab}
\end{equation}%
As a consequence of the boundary conditions (\ref{BCMIT}), one gets $n_{\mu }%
\bar{\psi}\gamma ^{\mu }\psi =0$ and the normal component of the fermionic
current vanishes on the edges. Here and in what follows, $\bar{\psi}=\psi
^{\dagger }\gamma ^{0}$ is the Dirac adjoint and the dagger denotes
Hermitian conjugation. This shows that the boundaries are impenetrable for
the fermionic field. The boundary condition of the type (\ref{BCMIT}) was
used in the bag model of hadrons for the confinement of quarks \cite{Chod74}%
. The analog boundary condition in graphene physics is referred as the
infinite mass boundary condition. It has been employed in (2+1)-dimensional
Dirac theory for the breaking of time-reversal symmetry without magnetic
fields \cite{Berr87}. Mainly motivated by the graphene physics, more general
conditions for the Dirac equation, ensuring the absence of current normal to
the boundary, have been discussed in \cite{Been08,Mcca04}. In
(2+1)-dimensions, the most general energy-independent boundary condition
contains four parameters.

For the further consideration we need to specify the representation for the
Dirac matrices. In Cartesian coordinates, we take $\gamma ^{(0)}=\sigma _{3}$%
, $\gamma ^{(l)}=i\sigma _{l}$, $l=1,2$, where $\sigma _{1}$, $\sigma _{2}$,
$\sigma _{3}$ are Pauli matrices. The Dirac matrices in cylindrical
coordinates are obtained in the standard way by using the corresponding
tetrad fields. They have the form $\gamma ^{0}=\sigma _{3}$ and
\begin{equation}
\gamma ^{l}=\frac{i^{2-l}}{r^{l-1}}\left(
\begin{array}{cc}
0 & e^{-i\phi } \\
(-1)^{l-1}e^{i\phi } & 0%
\end{array}%
\right) ,  \label{DirMat2}
\end{equation}%
for $l=1,2$. The corresponding spin connection vanishes, $\Gamma _{\mu }=0$.
For the vector potential we will consider a configuration corresponding to
the presence of a magnetic flux located in the region $r<a$. In the region
under consideration, $a\leqslant r\leqslant b$, for the covariant components
in the coordinates $(t,r,\phi )$ one has
\begin{equation}
A_{\mu }=(0,0,A_{2}).  \label{Amu}
\end{equation}%
Note that the physical azimuthal component for the vector potential is given
by $A_{\phi }=-A_{2}/r$ and for the magnetic flux threading the ring we have $%
\Phi =-2\pi A_{2}$. Though the magnetic field strength for (\ref{Amu}) vanishes,
the magnetic flux enclosed by the ring gives rise to Aharonov-Bohm-like
effects on physical observables, in particular for the VEVs. Note that the
distribution of the magnetic flux in the region $r<a$ can be arbitrary. As
the boundary $r=a$ is impenetrable for the fermionic field, the effect of
the gauge field is purely topological and depends on the total flux alone.
In this sense, the inner boundary can be viewed as a simplified model for a
finite radius magnetic flux with the reflecting wall.

The zero-point fluctuations of the fermionic field $\psi (x)$ in the region $%
a\leqslant r\leqslant b$ are influenced by the magnetic flux threading the
ring. As a consequence, the VEVs of physical quantities depend on the flux.
This gives rise to a number of interesting physical phenomena, such as
parity anomalies, formation of fermionic condensate and generation of
quantum numbers. Here we are interested in the VEV of the fermionic current $%
j^{\mu }=\bar{\psi}\gamma ^{\mu }\psi $. It appears as the source in the
semiclassical Maxwell equations and, hence, it plays an important role in
self-consistent dynamics involving the electromagnetic field. Let $%
S_{ik}^{(1)}(x,x^{\prime })=\langle 0|[\psi _{i}(x),\bar{\psi}_{k}(x^{\prime
})]|0\rangle $ be the fermion two-point function, where $i$ and $k$ are
spinor indices and $|0\rangle $ denotes the vacuum state. The VEV of the
current density is expressed in terms of this function as%
\begin{equation}
\langle j^{\mu }(x)\rangle \equiv \langle 0|j^{\mu }(x)|0\rangle =-\frac{e}{2%
}\mathrm{Tr}(\gamma ^{\mu }S^{(1)}(x,x)),  \label{VEVj}
\end{equation}%
with the trace over spinor indices understood. The expression in the
right-hand side can be presented in the form of the sum over a complete set
of positive- and negative-energy fermionic modes $\{\psi _{\sigma
}^{(+)}(x),\psi _{\sigma }^{(-)}(x)\}$, specified by a set of quantum
numbers $\sigma $. The mode functions $\psi _{\sigma }^{(\kappa )}(x)$, $%
\kappa =\pm $, obey the Dirac equation (\ref{Direq}) and the boundary
conditions (\ref{BCMIT}). Expanding the field operator in terms of $\psi
_{\sigma }^{(\kappa )}(x)$ and using the commutation relations for the
annihilation and creation operators, for the VEV of the current density the
following mode sum is obtained:
\begin{equation}
\langle j^{\mu }\rangle =-\frac{e}{2}\sum_{\sigma }\sum_{\kappa =-,+}\kappa
\bar{\psi}_{\sigma }^{(\kappa )}(x)\gamma ^{\mu }\psi _{\sigma }^{(\kappa
)}(x).  \label{VEVj1}
\end{equation}%
Our prescription is to find the complete set of modes and to use this mode
sum for the evaluation of the vacuum charge and current densities.

From the point of view of the renormalization of the VEVs, the important
point here is that, owing to the flatness of the background spacetime and
the zero field tensor for the external electromagnetic field in the region
under consideration, for points outside the boundaries the structure of
divergences is the same as for the (2+1)-dimensional boundary-free Minkowski
spacetime in the absence of the magnetic flux. Consequently, the
renormalization is reduced to the subtraction from the VEVs of the
corresponding Minkowskian quantity. In problems of the quantum field theory
with boundaries, one of the most efficient ways to extract from the VEVs the
boundary-free part in a regularization independent way is based on the
application of the Abel-Plana-type summation formulae to the corresponding
mode sums (for the applications of the Abel-Plana formula and its
generalizations in the theory of the Casimir effect see \cite%
{Most97,Saha08Book}).

\section{Fermionic modes}

\label{sec:Modes}

For the evaluation of the VEV in accordance with Eq. (\ref{VEVj1}) we need
to have the mode functions $\psi _{\sigma }^{(\pm )}(x)$ obeying the
boundary conditions (\ref{BCMIT}). Decomposing the spinor $\psi _{\sigma
}^{(\kappa )}(x)$, $\kappa =\pm $, into upper and lower components, $\varphi
_{+}$ and $\varphi _{-}$, respectively, from Eq. (\ref{Direq}) we get the
equations
\begin{equation}
\left( \partial _{0}\pm ism\right) \varphi _{\pm }\pm ie^{\mp i\phi }\left[
\partial _{1}\mp i(\partial _{2}+i\alpha )/r\right] \varphi _{\mp }=0,
\label{DirEqphi}
\end{equation}%
with the notation%
\begin{equation}
\alpha =eA_{2}=-\Phi /\Phi _{0},  \label{alfa}
\end{equation}%
where $\Phi _{0}=2\pi /e$ is the elementary flux or the flux quantum. From
Eq. (\ref{DirEqphi}) one finds the second-order differential equations for
the separate components:
\begin{equation}
\left[ \partial _{0}^{2}-\partial _{1}^{2}-\frac{1}{r}\partial _{1}-\frac{1}{%
r}(\partial _{2}+i\alpha )^{2}+m^{2}\right] \varphi _{\pm }=0.
\label{DireqPhi1}
\end{equation}%
Note that this equation is the same for both the components.

Let us present the solutions of these equations in the form%
\begin{equation}
\varphi _{\pm }=\chi _{\pm }(r)e^{-\kappa iEt+i\left( j_{\pm }-1/2\right)
\phi },  \label{phipl}
\end{equation}%
where $E>0$ and $j_{\pm }=\ldots ,-3/2,-1/2,1/2,3/2,\ldots $. For the
functions $\chi _{\pm }(r)$ we get the equations%
\begin{equation}
\left( -\kappa E\pm sm\right) \chi _{\pm }e^{i\left( j_{\pm }\pm 1-j_{\mp
}\right) \phi }\pm \left( \partial _{1}\pm \frac{j_{\pm }+\alpha -1/2}{r}%
\right) \chi _{\mp }=0.  \label{Eq1}
\end{equation}%
and%
\begin{equation}
\left[ \partial _{1}^{2}+\frac{1}{r}\partial _{1}+\gamma ^{2}-\frac{1}{r^{2}}%
\left( j_{\pm }+\alpha -1/2\right) ^{2}\right] \chi _{\pm }(r)=0,
\label{Eq2}
\end{equation}%
where $\gamma =\sqrt{E^{2}-m^{2}}$. From Eq. (\ref{Eq1}) it follows that $%
j_{-}=j_{+}+1$.

With this choice and denoting $j_{+}=j$, we take the solution for the
function with the upper sign as%
\begin{equation}
\chi _{+}(r)=Z_{\beta _{j}}(\gamma r)=c_{1}J_{\beta _{j}}(\gamma
r)+c_{2}Y_{\beta _{j}}(\gamma r),  \label{xip}
\end{equation}%
where $J_{\nu }(x)$ and $Y_{\nu }(x)$ are the Bessel and Neumann functions
and%
\begin{equation}
\beta _{j}=|j+\alpha |-\epsilon _{j}/2.  \label{betj}
\end{equation}%
Here, $\epsilon _{j}=1$ for $j>-\alpha $, $\epsilon _{j}=-1$ for $j\leqslant
-\alpha $. Note that $\epsilon _{j}\beta _{j}=j+\alpha -1/2$. By taking into
account the recurrence relations for the cylinder functions, from Eq. (\ref%
{Eq1}) it follows that
\begin{equation}
\chi _{-}(r)=\frac{\epsilon _{j}\gamma }{\kappa E+sm}Z_{\beta _{j}+\epsilon
_{j}}(\gamma r).  \label{xim}
\end{equation}

The ratio of the coefficients in the linear combination (\ref{xip}) is
determined from the boundary condition (\ref{BCMIT}) at $r=a$:
\begin{equation}
\frac{c_{2}}{c_{1}}=-\frac{J_{\beta _{j}}^{(a)}\left( \gamma a\right) }{%
Y_{\beta _{j}}^{(a)}\left( \gamma a\right) },  \label{c21}
\end{equation}%
where and in what follows, for the Bessel and Neumann functions, we use the
notation defined as%
\begin{eqnarray}
f_{\beta _{j}}^{(u)}\left( x\right) &=&xf_{\beta _{j}}^{\prime }\left(
x\right) +[n_{u}(\kappa \sqrt{x^{2}+m_{u}^{2}}+sm_{u})-\epsilon _{j}\beta
_{j}]f_{\beta _{j}}\left( x\right)  \notag \\
&=&n_{u}(\kappa \sqrt{x^{2}+m_{u}^{2}}+sm_{u})f_{\beta _{j}}\left( x\right)
-\epsilon _{j}xf_{\beta _{j}+\epsilon _{j}}\left( x\right) ,  \label{fbar}
\end{eqnarray}%
with $u=a,b$, $f=J,Y$, and $m_{u}=mu$. As a consequence, the mode functions
are written in the form%
\begin{equation}
\psi _{\sigma }^{(\kappa )}(x)=C_{\kappa }e^{-\kappa iEt+ij\phi }\left(
\begin{array}{c}
g_{\beta _{j},\beta _{j}}\left( \gamma a,\gamma r\right) e^{-i\phi /2} \\
\frac{\epsilon _{j}\gamma e^{i\phi /2}}{\kappa E+sm}g_{\beta _{j},\beta
_{j}+\epsilon _{j}}\left( \gamma a,\gamma r\right)%
\end{array}%
\right) ,  \label{psip}
\end{equation}%
where%
\begin{equation}
g_{\beta _{j},\mu }(x,y)=Y_{\beta _{j}}^{(a)}\left( x\right) J_{\mu }\left(
y\right) -J_{\beta _{j}}^{(a)}\left( x\right) Y_{\mu }\left( y\right) .
\label{ge}
\end{equation}%
We can check that the modes (\ref{psip}) are eigenfunctions of the total
angular momentum operator $\hat{J}=-i(\partial _{2}+ieA)+\sigma _{3}/2$, for
the eigenvalues $j+\alpha $:%
\begin{equation}
\hat{J}\psi _{\sigma }^{(\kappa )}(x)=(j+\alpha )\psi _{\sigma }^{(\kappa
)}(x).  \label{J}
\end{equation}%
Here, the part $\sigma _{3}/2$ corresponds to the pseudospin.

In a similar way, from the boundary condition (\ref{BCMIT}) at $r=b$ we get $%
c_{2}/c_{1}=-J_{\beta _{j}}^{(b)}\left( \gamma b\right) /Y_{\beta
_{j}}^{(b)}\left( \gamma b\right) $. Combining this with Eq. (\ref{c21}),
one concludes that the eigenvalues for $\gamma $ in the region $a\leqslant
r\leqslant b$ are the roots of the equation%
\begin{equation}
C_{\beta _{j}}(\eta ,\gamma a)\equiv J_{\beta _{j}}^{(a)}\left( \gamma
a\right) Y_{\beta _{j}}^{(b)}\left( \gamma b\right) -J_{\beta
_{j}}^{(b)}\left( \gamma b\right) Y_{\beta _{j}}^{(a)}\left( \gamma a\right)
=0,  \label{EigEq}
\end{equation}%
with $\eta =b/a$. The positive solutions of this equation with respect to $%
\gamma a$ will be denoted by $z_{l}$, $l=1,2,\ldots $, $z_{l}<z_{l+1}$. For
the eigenvalues of $\gamma $ one has $\gamma =\gamma _{l}=z_{l}/a$. In this
way, the mode functions are specified by the quantum numbers $\sigma =(l,j)$%
. Note that, the roots $z_{l}$ depend on the value of $j$ as well. In order
to simplify the expressions below we do not write this dependence
explicitly. For a given $j$, the equations (\ref{EigEq}) for the eigenvalues
$\gamma _{l}$ of the positive- and negative-energy modes differ by the
change of the energy sign, $E\rightarrow -E$ (through $\kappa $ in Eq. (\ref%
{fbar})). For the energy one has $E=\sqrt{\gamma _{l}^{2}+m^{2}}$. As we
see, in the case of a massless field, the finite size effects induce a gap
in the energy spectrum. The gap can be controlled by the geometrical
characteristics of the model. Note that the gap generated by the finite size
effects plays an important role in graphene made nanoribbons. In the problem
at hand the size of the energy gap is determined by the minimal value of $%
\gamma _{l}$. For example, in the case $\alpha =1/3$, $j=1/2$ for the first
root one has $z_{1}\approx 1.59$ for $b/a=2$ and $z_{1}\approx 0.81$ for $%
b/a=3$. The root increases with decreasing $b/a$ and with increasing $j$.

The function $C_{\beta _{j}}(\eta ,x)$ in Eq. (\ref{EigEq}) can also be
written in terms of the Hankel functions $H_{\beta _{j}}^{(1,2)}(x)$ as%
\begin{equation}
C_{\nu }(\eta ,x)=(i/2)\left[ H_{\beta _{j}}^{(2a)}\left( x\right) H_{\beta
_{j}}^{(1b)}\left( \eta x\right) -H_{\beta _{j}}^{(1a)}\left( x\right)
H_{\beta _{j}}^{(2b)}\left( \eta x\right) \right] ,  \label{CHank}
\end{equation}%
with the notations defined as Eq. (\ref{fbar}). With this representation, we
can see that for a massless field the equation $C_{\beta _{j}}(\eta ,\gamma
a)=0$ is reduced to the one given in \cite{Rech07} for graphene rings
described by the Dirac model with the infinite mass boundary condition on
the edges. Note that to obtain an analytical approximation of the spectrum,
in \cite{Rech07} the asymptotic form of the Hankel functions for large
arguments was used. This approximation is valid for rings with the radius
much larger than the width. As it will be shown below, the use of the
generalized Abel-Plana formula allows us to obtain closed analytic
expressions for the VEVs in the general case of geometrical characteristics
of the ring.

For the complete specification of the mode functions (\ref{psip}) it remains
to determine the coefficient $C_{\kappa }$. The latter is found from the
orthonormalization condition
\begin{equation}
\int_{a}^{b}dr\int_{0}^{2\pi }d\phi \,r\psi _{\sigma }^{(\kappa )\dagger
}(x)\psi _{\sigma ^{\prime }}^{(\kappa )}(x)=\delta _{jj^{\prime }}\ \delta
_{ll^{\prime }}.  \label{Norm}
\end{equation}%
Substituting the modes, using the result for the integral involving the
square of cylinder functions (see \cite{Prud86}), after long but
straightforward calculations one finds%
\begin{equation}
\left\vert C_{\kappa }\right\vert ^{2}=\frac{\pi z_{l}}{16a^{2}}\frac{%
E+\kappa sm}{E}T_{\beta _{j}}^{ab}(\eta ,z_{l}).  \label{C+}
\end{equation}%
Here, we have introduced the notation%
\begin{equation}
T_{\beta _{j}}^{ab}(\eta ,z)=z\left[ D_{b}J_{\beta _{j}}^{(a)2}\left(
z\right) /J_{\beta _{j}}^{(b)2}\left( \eta z\right) -D_{a}\right] ^{-1},
\label{Tab}
\end{equation}%
with%
\begin{equation}
D_{u}=u^{2}\frac{E+\kappa sm}{E}\left[ E\left( E+\kappa n_{u}\frac{\epsilon
_{j}\beta _{j}}{u}\right) +\kappa n_{u}\frac{E-\kappa sm}{2u}\right] .
\label{Du}
\end{equation}

As it has been already mentioned, the eigenvalue equations (\ref{EigEq}) for
the positive- and negative-energy modes are obtained from each other by the
change of the energy sign. Redefining the azimuthal quantum number in
accordance with $j\rightarrow -j$, the negative-energy modes can also be
written in the form%
\begin{equation}
\psi _{\sigma }^{(-)}=C_{-}^{\prime }e^{iEt-ij\phi }\left(
\begin{array}{c}
\frac{\epsilon _{j}^{-}\gamma e^{-i\phi /2}}{E+sm}g_{\beta _{j}^{-},\beta
_{j}^{-}+\epsilon _{\lambda _{n}^{-}}}(\gamma a,\gamma r) \\
g_{\beta _{j}^{-},\beta _{j}^{-}}(\gamma a,\gamma r)e^{i\phi /2}%
\end{array}%
\right) ,  \label{psim}
\end{equation}%
where $\beta _{j}^{-}=|j-\alpha |-\epsilon _{j}^{-}/2$, $\epsilon _{j}^{-}=1$
for $j>\alpha $ and $\epsilon _{j}^{-}=-1$ for $j\leqslant \alpha $. In the
definition (\ref{ge}) of the function $g_{\beta _{j}^{-},\mu }(x,y)$ the
notations $f_{\beta _{j}^{-}}^{(u)}\left( x\right) $ are defined as Eq. (\ref%
{fbar}) with $\kappa =1$ (as in the case of the positive-frequency
functions). With the modes (\ref{psim}), the eigenvalues for $\gamma $ are
solutions of the equation%
\begin{equation}
J_{\beta _{j}^{-}}^{(a)}\left( \gamma a\right) Y_{\beta
_{j}^{-}}^{(b)}\left( \gamma b\right) -Y_{\beta _{j}^{-}}^{(a)}\left( \gamma
a\right) J_{\beta _{j}^{-}}^{(b)}\left( \gamma b\right) =0.  \label{EigEqm}
\end{equation}%
Because now the functions $f_{\nu }^{(u)}\left( x\right) $ are the same for
the positive- and negative-energy modes, Eq. (\ref{EigEqm}) differs from the
corresponding equation (\ref{EigEq}) for the positive-energy modes just by
the sign of $\alpha $, $\alpha \rightarrow -\alpha $. It can be seen that
the normalization constant $C_{-}^{\prime }$ is given by the expression in
the right-hand side of Eq. (\ref{C+}) with $\kappa =1$ and with $%
z_{l}=\gamma a$ being the roots of Eq. (\ref{EigEqm}). Of course, in the
evaluation of the VEVs we can use both types of the modes.

\section{Charge density}

\label{sec:Charge}

We start our consideration of the VEVs with the charge density. Substituting
the mode functions (\ref{psip}) into the mode-sum formula (\ref{VEVj1}) we
get%
\begin{equation}
\langle j^{0}\rangle =-\frac{\pi e}{32a^{2}}\sum_{j}\sum_{l=1}^{\infty
}\sum_{\kappa =\pm }\,\kappa T_{\beta _{j}}^{ab}(\eta ,z_{l})h_{0}(z_{l}),
\label{j0}
\end{equation}%
where $\sum_{j}$ stands for the summation over $j=\pm 1/2,\pm 3/2,\ldots $,
and%
\begin{equation}
h_{0}(z)=\frac{z}{E}\left[ \left( E+\kappa sm\right) g_{\beta _{j},\beta
_{j}}^{2}(z,zr/a)+\left( E-\kappa sm\right) g_{\beta _{j},\beta
_{j}+\epsilon _{j}}^{2}(z,zr/a)\right] ,  \label{h0}
\end{equation}%
with $E=\sqrt{z^{2}/a^{2}+m^{2}}$. In Eq. (\ref{j0}), the eigenvalues $z_{l}$
are given implicitly, as roots of Eq. (\ref{EigEq}), and this representation
is not convenient for the further evaluation of the VEV. Another
disadvantage of the representation (\ref{j0}) is that the separate terms in
the series are highly oscillating for large values of the quantum numbers.

These difficulties are overcome by using for the summation over $l$ the
Abel-Plana-type formula%
\begin{eqnarray}
&&\sum_{l=1}^{\infty }h(z_{l})T_{\beta _{j}}^{ab}(\eta ,z_{l})=\frac{4}{\pi
^{2}}\int_{0}^{\infty }{dz\,}\frac{h(z)}{J_{\beta _{j}}^{(a)2}\left(
z\right) +Y_{\beta _{j}}^{(a)2}\left( z\right) }  \notag \\
&&\qquad -\frac{2}{\pi }\underset{z=0}{\mathrm{Res}}\left[ \frac{%
h(z)H_{\beta _{j}}^{(1b)}\left( \eta z\right) }{C_{\beta _{j}}(\eta
,z)H_{\beta _{j}}^{(1a)}\left( z\right) }\right] -\frac{1}{\pi }%
\int_{0}^{\infty }dz\,\sum_{p=+,-}\Omega _{a\beta _{j}}^{(p)}(z,\eta
z)h(ze^{pi\frac{\pi }{2}}),  \label{Sum}
\end{eqnarray}%
where
\begin{equation}
\Omega _{a\beta _{j}}^{(p)}(z,\eta z)=\frac{K_{\beta _{j}}^{(bp)}(\eta
z)/K_{\beta _{j}}^{(ap)}(z)}{K_{\beta _{j}}^{(ap)}\left( z\right) I_{\beta
_{j}}^{(bp)}\left( \eta z\right) -I_{\beta _{j}}^{(ap)}(z)K_{\beta
_{j}}^{(bp)}(\eta z)}.  \label{Om}
\end{equation}%
Here we have introduced the notation%
\begin{equation}
f_{\beta _{j}}^{(up)}(x)=xf_{\beta _{j}}^{\prime }\left( x\right) +\left\{
n_{u}\left[ \kappa \sqrt{\left( xe^{p\pi i/2}\right) ^{2}+m_{u}^{2}}+sm_{u}%
\right] -\epsilon _{j}\beta _{j}\right\} f_{\beta _{j}}\left( x\right) ,
\label{fjp}
\end{equation}%
with $f=I,K$ for the modified Bessel functions $I_{\beta _{j}}\left(
x\right) $ and $K_{\beta _{j}}\left( x\right) $. The formula (\ref{Sum}) is
valid for a function $h(z)$ analytic in the complex half-plane Re$\,z>0$, $%
z=x+iy$, and obeying the condition $|h(z)|<\varepsilon (x)e^{c|y|}$, where $%
c<2(\eta -1)$ and $\varepsilon (x)/x\rightarrow 0$ for $x\rightarrow +\infty
$. On the imaginary axis the function $h(z)$ may have branch points. The
summation formula (\ref{Sum}) is obtained from the generalized Abel-Plana
formula \cite{Saha08Book} (see also \cite{Beze06}). Note that for the square
root in Eq. (\ref{fjp}) one has%
\begin{equation}
\sqrt{\left( ze^{p\pi i/2}\right) ^{2}+m_{u}^{2}}=\left\{
\begin{array}{cc}
\sqrt{m_{u}^{2}-z^{2}}, & z<m_{u} \\
pi\sqrt{z^{2}-m_{u}^{2}}, & z>m_{u}%
\end{array}%
\right. .  \label{Square}
\end{equation}%
In particular, we see that $f_{\beta _{j}}^{(u+)}(z)=f_{\beta
_{j}}^{(u-)}(z) $ for $z<m_{u}$. From here it follows that $\Omega _{a\beta
_{j}}^{(-)}(z,\eta z)=\Omega _{a\beta _{j}}^{(+)}(z,\eta z)$ for $z<ma$.

For the charge density the function $h(z)$ in the summation formula (\ref%
{Sum}) is specified by Eq. (\ref{h0}). This function has branch points $%
z=\pm ima$ on the imaginary axis. For $z<ma$ one has the relation $%
h_{0}(ze^{-\pi i/2})=-h_{0}(ze^{\pi i/2})$ and, hence, in the last integral
of Eq. (\ref{Sum}) the part over the region $[0,ma]$ becomes zero. For $z>ma$
we find%
\begin{equation}
h_{0}(ze^{pi\frac{\pi }{2}})=\frac{4z}{\pi ^{2}}\left[ \left( \frac{\kappa sm%
}{\sqrt{z^{2}/a^{2}-m^{2}}}+pi\right) G_{\beta _{j},\beta
_{j}}^{(ap)2}(z,zr/a)+\left( \frac{\kappa sm}{\sqrt{z^{2}/a^{2}-m^{2}}}%
-pi\right) G_{\beta _{j},\beta _{j}+\epsilon _{j}}^{(ap)2}(z,zr/a)\right] ,
\label{h0i}
\end{equation}%
with the function%
\begin{equation}
G_{\beta _{j},\mu }^{(ap)}(x,y)=K_{\beta _{j}}^{(ap)}\left( x\right) I_{\mu
}\left( y\right) -(-1)^{\mu -\beta _{j}}I_{\beta _{j}}^{(ap)}\left( x\right)
K_{\mu }\left( y\right) .  \label{Gp}
\end{equation}%
Substituting Eq. (\ref{h0i}) into Eq. (\ref{Sum}) and then into Eq. (\ref{j0}%
), we can see that $\kappa $ and $p$ enter into the expression of the part
corresponding to the last term in the right-hand side of Eq. (\ref{Sum}) in
the form of the product $\kappa p$. From here it follows that the positive-
and negative-energy modes give the same contributions to this part of the
charge density.

After all these transformations, the VEV of the charge density is presented
in the form%
\begin{eqnarray}
\langle j^{0}\rangle &=&\langle j^{0}\rangle _{a}+\frac{e}{2\pi ^{2}}%
\sum_{j}\,\int_{m}^{\infty }dx\,x\left\{ \frac{sm}{\sqrt{x^{2}-m^{2}}}\right.
\notag \\
&&\times \mathrm{Re}\left[ \Omega _{a\beta _{j}}(ax,bx)\left( G_{\beta
_{j},\beta _{j}}^{(a)2}(ax,rx)+G_{\beta _{j},\beta _{j}+\epsilon
_{j}}^{(a)2}(ax,rx)\right) \right]  \notag \\
&&\left. -\mathrm{Im}\left[ \Omega _{a\beta _{j}}(ax,bx)\left( G_{\beta
_{j},\beta _{j}}^{(a)2}(ax,rx)-G_{\beta _{j},\beta _{j}+\epsilon
_{j}}^{(a)2}(ax,rx)\right) \right] \right\} ,  \label{j02}
\end{eqnarray}%
where
\begin{equation}
\langle j^{0}\rangle _{a}=-\frac{e}{8\pi a^{2}}\sum_{j}\sum_{\kappa =\pm
1}\,\int_{0}^{\infty }{dz\,}\frac{\kappa h_{0}(z)}{J_{\beta
_{j}}^{(a)2}\left( z\right) +Y_{\beta _{j}}^{(a)2}\left( z\right) }.
\label{j0a}
\end{equation}%
The new notations are defined as%
\begin{equation}
\Omega _{a\beta _{j}}(ax,bx)=\frac{K_{\beta _{j}}^{(b)}(bx)/K_{\beta
_{j}}^{(a)}(ax)}{K_{\beta _{j}}^{(a)}\left( ax\right) I_{\beta
_{j}}^{(b)}\left( bx\right) -I_{\beta _{j}}^{(a)}(ax)K_{\beta _{j}}^{(b)}(bx)%
},  \label{Oma}
\end{equation}%
and
\begin{equation}
G_{\beta _{j},\mu }^{(u)}(x,y)=K_{\beta _{j}}^{(u)}\left( x\right) I_{\mu
}\left( y\right) -(-1)^{\mu -\beta _{j}}I_{\beta _{j}}^{(u)}\left( x\right)
K_{\mu }\left( y\right) ,  \label{Gen}
\end{equation}%
with $u=a,b$ (the function $G_{\beta _{j},\mu }^{(b)}(x,y)$ is used below).
For the modified Bessel functions now we use the notations%
\begin{eqnarray}
f_{\beta _{j}}^{(u)}(z) &=&zf_{\beta _{j}}^{\prime }\left( z\right) +\left[
n_{u}\left( i\sqrt{z^{2}-m_{u}^{2}}+sm_{u}\right) -\epsilon _{j}\beta _{j}%
\right] f_{\beta _{j}}\left( z\right)  \notag \\
&=&\delta _{f}zf_{\beta _{j}+\epsilon _{j}}\left( z\right) +n_{u}(i\sqrt{%
z^{2}-m_{u}^{2}}+sm_{u})f_{\beta _{j}}\left( z\right) ,  \label{fnu}
\end{eqnarray}%
where $f=I,K$, $\delta _{I}=1$ and $\delta _{K}=-1$.

Let us present the parameter $\alpha $ from Eq. (\ref{alfa}), related to the
magnetic flux threading the ring, in the form%
\begin{equation}
\alpha =N+\alpha _{0},\;|\alpha _{0}|\leqslant 1/2,  \label{alfdec}
\end{equation}%
where $N$ is an integer. Redefining the summation variable $j$ in accordance
with $j+N\rightarrow j$ we see that the charge density does not depend on
the integer part $N$. Then, separating the summations over negative and
positive values of $j$, making the replacement $j\rightarrow -j$ in the part
with negative $j$ and introducing a new summation variable $n=j-1/2$, the
charge density is presented as%
\begin{eqnarray}
\langle j^{0}\rangle &=&\langle j^{0}\rangle _{a}+\frac{e}{2\pi ^{2}}%
\sum_{n=0}^{\infty }\,\sum_{p=\pm }p\int_{m}^{\infty }dx\,x\left\{ \frac{sm}{%
\sqrt{x^{2}-m^{2}}}\right.  \notag \\
&&\times \mathrm{Re}\left[ \Omega _{an_{p}}(ax,bx)\left(
G_{n_{p},n_{p}}^{(a)2}(ax,rx)+G_{n_{p},n_{p}+1}^{(a)2}(ax,rx)\right) \right]
\notag \\
&&\left. +\mathrm{Im}\left[ \Omega _{an_{p}}(ax,bx)\left(
G_{n_{p},n_{p}+1}^{(a)2}(ax,rx)-G_{n_{p},n_{p}}^{(a)2}(ax,rx)\right) \right]
\right\} ,  \label{j02bn}
\end{eqnarray}%
with%
\begin{equation}
n_{p}=n+p\alpha _{0},  \label{np}
\end{equation}%
and now the notation (\ref{fnu}) in the definitions (\ref{Oma}) and (\ref%
{Gen}) for $\Omega _{an_{p}}(ax,bx)$ and $G_{n_{p},\mu }^{(a)}(ax,rx)$ is
specified to%
\begin{equation}
f_{n_{p}}^{(u)}(z)=\delta _{f}zf_{n_{p}+1}\left( z\right) +n_{u}(sm_{u}+i%
\sqrt{z^{2}-m_{u}^{2}})f_{n_{p}}\left( z\right) ,  \label{fnun}
\end{equation}%
for $f=I,K$. The representation (\ref{j02bn}) explicitly shows that the last
term is an odd function of the fractional part $\alpha _{0}$. The property
that the VEVs do not depend on the integer part of the flux, in units of the
flux quantum, is a general feature in the Aharonov-Bohm effect and is a
consequence of that the flux enters through the phase of the wavefunction.

The last term in Eq. (\ref{j02bn}) vanishes in the limit $b\rightarrow
\infty $ (for large values $b$ the function $\Omega _{an_{p}}(ax,bx)$ falls
off as $e^{-2bx}$). From here it follows that the part (\ref{j0a}) is the
charge density in the region $r\geqslant a$ for the geometry of a single
boundary at $r=a$. In order to extract from that part the boundary-induced
effects we further transform the expression (\ref{j0a}), with $h_{0}(z)$
form (\ref{h0}), by using the relation%
\begin{equation}
\frac{g_{\beta _{j},\lambda }^{2}(z,y)}{J_{\beta _{j}}^{(a)2}(z)+Y_{\beta
_{j}}^{(a)2}(z)}=J_{\lambda }^{2}(y)-\frac{1}{2}\sum_{l=1,2}\frac{J_{\beta
_{j}}^{(a)}(z)}{H_{\beta _{j}}^{(la)}(z)}H_{\lambda }^{(l)2}(y),
\label{Rel1}
\end{equation}%
with $\lambda =\beta _{j},\beta _{j}+\epsilon _{j}$. Substituting into Eq. (%
\ref{j0a}), in the part with the Hankel functions, we rotate the contour of
the integration over $z$ by the angles $\pi /2$ and $-\pi /2$ for the terms
with $l=1$ and $l=2$, respectively. The integrals over the segments $%
[0,im_{a}]$ and $[0,-im_{a}]$ cancel each other and, introducing the
modified Bessel functions, again, we can see that the contributions coming
from the positive- and negative-energy modes coincide. As a result, for the
contribution (\ref{j0a}), we come to the decomposition%
\begin{equation}
\langle j^{0}\rangle _{a}=\langle j^{0}\rangle _{0}+\langle j^{0}\rangle
_{a}^{\mathrm{(b)}},  \label{j0adec}
\end{equation}%
where%
\begin{equation}
\langle j^{0}\rangle _{0}=-\frac{e}{8\pi }\sum_{j}\sum_{\kappa =\pm
}\,\kappa \int_{0}^{\infty }dx\,x\left[ \left( 1+\frac{s\kappa m}{\sqrt{%
x^{2}+m^{2}}}\right) J_{\beta _{j}}^{2}(xr)+\left( 1-\frac{s\kappa m}{\sqrt{%
x^{2}+m^{2}}}\right) J_{\beta _{j}+\epsilon _{j}}^{2}(xr)\right] ,
\label{j00}
\end{equation}%
and%
\begin{eqnarray}
\langle j^{0}\rangle _{a}^{\mathrm{(b)}} &=&\frac{e}{2\pi ^{2}}%
\sum_{j}\,\int_{m}^{\infty }dx\,x\left\{ sm\frac{K_{\beta
_{j}}^{2}(rx)+K_{\beta _{j}+\epsilon _{j}}^{2}(rx)}{\sqrt{x^{2}-m^{2}}}%
\mathrm{Re}\left[ \frac{I_{\beta _{j}}^{(a)}(ax)}{K_{\beta _{j}}^{(a)}(ax)}%
\right] \right.  \notag \\
&&\left. +\left[ K_{\beta _{j}+\epsilon _{j}}^{2}(rx)-K_{\beta _{j}}^{2}(rx)%
\right] \mathrm{Im}\left[ \frac{I_{\beta _{j}}^{(a)}(ax)}{K_{\beta
_{j}}^{(a)}(ax)}\right] \right\} ,  \label{j0a2}
\end{eqnarray}%
with the notations (\ref{fnu}). For the representation $s=1$, this
expression for a single boundary-induced part coincides with the one given
in Ref. \cite{Beze10} (the sign difference is related to that in \cite%
{Beze10}, for the evaluation of the VEVs, the analog of the negative-energy
mode functions (\ref{psim}) for the geometry with a single boundary was used
with $\alpha $ replaced by $-\alpha $; hence, in comparing the formulas here
with the results of \cite{Beze10}, the replacements $\alpha \rightarrow
-\alpha $ and $\alpha _{0}\rightarrow -\alpha _{0}$ should be made).

In order to give a physical interpretation of the separate terms in Eq. (\ref%
{j0adec}) let us consider the limit $a\rightarrow 0$. Note that the radius
of the magnetic flux should also be taken to zero. For $j+\alpha \neq 0$ one
has%
\begin{equation}
\frac{I_{\beta _{j}}^{(a)}(ax)}{K_{\beta _{j}}^{(a)}(ax)}\approx \frac{i%
\sqrt{x^{2}-m^{2}}+\epsilon _{j}sm}{x}\frac{2\left( ax/2\right) ^{2|j+\alpha
|}}{\Gamma ^{2}(|j+\alpha |+1/2)},  \label{IKrata0}
\end{equation}%
and, hence, the part (\ref{j0a2}) vanishes as $a^{2|j+\alpha |}$. For
half-odd integer values of $\alpha $, the exceptional case corresponds to
the mode with $j=-\alpha $. For this mode $\beta _{j}=1/2$, $\epsilon
_{j}=-1 $, and we get
\begin{equation}
\frac{I_{\beta _{j}}^{(a)}(ax)}{K_{\beta _{j}}^{(a)}(ax)}\approx \frac{2}{%
\pi }\frac{i\sqrt{x^{2}-m^{2}}-sm}{x}.  \label{IKrata01}
\end{equation}%
Note that for this special value of $\beta _{j}$, in Eq. (\ref{j0a2}) the
coefficient of the term with the imaginary part of Eq. (\ref{IKrata01})
vanishes.

Hence, if $\alpha $ is not a half-odd integer one has%
\begin{equation}
\lim_{a\rightarrow 0}\langle j^{0}\rangle _{a}=\langle j^{0}\rangle _{0}.
\label{lima1}
\end{equation}%
In this case, the part (\ref{j0a2}) in the VEV of the charge density is
induced by the presence of the boundary at $r=a$, whereas $\langle
j^{0}\rangle _{0}$ gives the charge density in the boundary-free geometry
with a point like magnetic flux at $r=0$. For $\alpha $ being a half-odd
integer, by using (\ref{IKrata01}), from Eq. (\ref{j0a2}) one gets%
\begin{equation}
\lim_{a\rightarrow 0}\langle j^{0}\rangle _{a}=\langle j^{0}\rangle _{0}+%
\frac{em}{\pi ^{2}r}\,\int_{0}^{\infty }dy\,\frac{e^{-2mr\sqrt{y^{2}+1}}}{%
y^{2}+1}.  \label{lima2}
\end{equation}%
For a massless field the last term in this expression vanishes and we come
to the same interpretation of the separate terms in Eq. (\ref{j0adec}).

The boundary-induced contribution (\ref{j0a2}) does not depend on the
integer part $N$ in Eq. (\ref{alfdec}). Redefining the summation variable in
Eq. (\ref{j0a2}), this contribution is rewritten in the form
\begin{eqnarray}
\langle j^{0}\rangle _{a}^{\mathrm{(b)}} &=&\frac{e}{2\pi ^{2}}%
\sum_{n=0}^{\infty }\,\sum_{p=-,+}p\int_{m}^{\infty }dx\,x  \notag \\
&&\times \left\{ sm\frac{K_{n_{p}}^{2}(rx)+K_{n_{p}+1}^{2}(rx)}{\sqrt{%
x^{2}-m^{2}}}\mathrm{Re}\left[ \frac{I_{n_{p}}^{(a)}(ax)}{K_{n_{p}}^{(a)}(ax)%
}\right] \right.  \notag \\
&&\left. +\left[ K_{n_{p}+1}^{2}(rx)-K_{n_{p}}^{2}(rx)\right] \mathrm{Im}%
\left[ \frac{I_{n_{p}}^{(a)}(ax)}{K_{n_{p}}^{(a)}(ax)}\right] \right\} ,
\label{j0np}
\end{eqnarray}%
with $n_{p}$ given by Eq. (\ref{np}). This explicitly shows that the single
boundary-induced charge density is an odd function of $\alpha _{0}$. The
real and imaginary parts in Eq. (\ref{j0np}) are explicitly given by the
relation%
\begin{equation}
\frac{I_{n_{p}}^{(u)}(z)}{K_{n_{p}}^{(u)}(z)}=\frac{W_{n_{p}}^{(u)}(z)-in_{u}%
\sqrt{1-m_{u}^{2}/z^{2}}}{z[K_{n_{p}+1}^{2}\left( z\right)
+K_{n_{p}}^{2}\left( z\right) ]-2sn_{u}m_{u}K_{n_{p}}\left( z\right)
K_{n_{p}+1}\left( z\right) },  \label{IK}
\end{equation}%
with $u=a,b$ and with the function%
\begin{eqnarray}
W_{\nu }^{(u)}(z) &=&z\left[ I_{\nu }\left( z\right) K_{\nu }\left( z\right)
-I_{\nu +1}\left( z\right) K_{\nu +1}\left( z\right) \right]  \notag \\
&&+n_{u}sm_{u}\left[ I_{\nu +1}\left( z\right) K_{\nu }\left( z\right)
-I_{\nu }\left( z\right) K_{\nu +1}\left( z\right) \right] .  \label{W}
\end{eqnarray}%
Note that for $z\geqslant m_{u}$ the denominator in Eq. (\ref{IK}) is
positive. For a massless field and at large distances from the boundary, $%
r\gg a$, the dominant contribution to the boundary-induced part (\ref{j0np})
comes from the term $n=0$ and this part decays as $(a/r)^{3-2|\alpha _{0}|}$
with the sign $\mathrm{sgn}(\alpha _{0})\langle j^{0}\rangle _{a}^{\mathrm{%
(b)}}/e<0$. For a massive field and for $r\gg a,m^{-1}$, the dominant
contribution in Eq. (\ref{j0np}) comes from the region near the lower limit
of the integration and the boundary-induced charge density is suppressed by
the factor $e^{-2mr}/r^{3/2}$.

From the consideration above it follows that, if $|\alpha _{0}|\neq 1/2$,
the part $\langle j^{0}\rangle _{0}$ can be interpreted as the charge
density in boundary-free two-dimensional space with a special type of
boundary condition on the magnetic flux line at $r=0$. Namely, we impose the
bag boundary condition at finite radius which is then taken to zero.
Consequently, the part (\ref{j0np}) is interpreted as the contribution
induced in the region $a\leqslant r<\infty $ by the boundary $r=a$. The last
term in Eq. (\ref{j02bn}) is the contribution in the charge density induced
when we add the boundary at $r=b$ to the geometry with a single boundary at $%
r=a$. In this sense, this part can be termed as the second boundary-induced
contribution.

The expression (\ref{j00}) for the boundary-free part can be further
simplified. The first terms in the brackets of the coefficients of the
functions $J_{\beta _{j}}^{2}(xr)$ and $J_{\beta _{j}+\epsilon _{j}}^{2}(xr)$
are canceled for the contributions coming from the positive- and
negative-energy modes. For the remaining part we get%
\begin{equation}
\langle j^{0}\rangle _{0}=\frac{esm}{4\pi }\sum_{j}\,\int_{0}^{\infty }dx\,x%
\frac{J_{\beta _{j}+\epsilon _{j}}^{2}(xr)-J_{\beta _{j}}^{2}(xr)}{\sqrt{%
x^{2}+m^{2}}},  \label{j002}
\end{equation}%
that is further simplified to%
\begin{equation}
\langle j^{0}\rangle _{0}=\frac{esm}{4\pi }\,\int_{0}^{\infty }dx\,x\frac{%
J_{-\alpha _{0}}^{2}(xr)-J_{\alpha _{0}}^{2}(xr)}{\sqrt{x^{2}+m^{2}}}.
\label{j003}
\end{equation}%
This expression with $s=1$ was obtained in \cite{Jaro86,Flek91}. After the
rotation of the integration contour, it can also be presented in the form
\cite{Jaro86,Flek91}
\begin{equation}
\langle j^{0}\rangle _{0}=\frac{esm^{2}}{\pi ^{3}}\sin \left( \pi \alpha
_{0}\right) \int_{1}^{\infty }dx\frac{xK_{\alpha _{0}}^{2}(mrx)}{\sqrt{%
x^{2}-1}}.  \label{j002d}
\end{equation}%
An equivalent expression for the boundary-free part is provided in Ref. \cite%
{Beze10}:%
\begin{equation}
\langle j^{0}\rangle _{0}=\frac{esm}{2\pi ^{2}r}\sin \left( \pi \alpha
_{0}\right) \int_{0}^{\infty }dx\,\frac{\cosh (2\alpha _{0}x)}{\cosh x}%
e^{-2mr\cosh x}.  \label{j002c}
\end{equation}%
Similarly to the boundary-induced contributions, this part is an odd
function of the parameter $\alpha _{0}$. For a massless field the
boundary-free contribution in the charge density vanishes for $r\neq 0$. In
the case of a massive field, at large distances, the charge density $\langle
j^{0}\rangle _{0}$ falls off as $e^{-2mr}/r^{3/2}$, whereas at the origin it
diverges as $1/r$. At large distances, the decaying factor for a massive
field is the same as that for the boundary-induced contribution (\ref{j0np}).

The charge density (\ref{j002d}) for the boundary-free geometry with a point
like magnetic flux corresponds to a special boundary condition on the
fermion field at the location of the flux. In general, the self-adjoint
extension procedure for the Dirac Hamiltonian leads to a one-parameter
family of boundary conditions \cite{Sous89}. The value of the parameter in
the boundary condition is determined by the physical details of the magnetic
field distribution inside a more realistic finite radius flux tube (see,
e.g., \cite{Hage90} for models with finite radius magnetic flux).

Combining all the results given above, we conclude that the total charge
density $\langle j^{0}\rangle $ in the region $a\leqslant r\leqslant b$ is a
periodic odd function of the magnetic flux threading the ring, with the
period equal to the flux quantum. As a function of the parameter $\alpha $
(magnetic flux in units of the flux quantum), the charge density (\ref{j002c}%
) in the boundary-free geometry is discontinuous at the half-odd integer
values $\alpha =N+1/2$:%
\begin{equation}
\lim_{\alpha _{0}\rightarrow \pm 1/2}\langle j^{0}\rangle _{0}=\pm \frac{esm%
}{2\pi ^{2}r}K_{0}(2mr).  \label{j00half}
\end{equation}%
In the geometry with a single boundary at $r=a$, it can be seen that for the
boundary-induced contribution in the region $r\geqslant a$ one has $%
\lim_{\alpha _{0}\rightarrow \pm 1/2}\langle j^{0}\rangle _{a}^{\mathrm{(b)}%
}=-\lim_{\alpha _{0}\rightarrow \pm 1/2}\langle j^{0}\rangle _{0}$. This
means that, in this geometry, the total charge density vanishes for $\alpha
=N+1/2$, $\lim_{\alpha _{0}\rightarrow \pm 1/2}\langle j^{0}\rangle _{a}=0$,
and it is a continuous function of the magnetic flux everywhere. It can be
checked that, in the expression (\ref{j02bn}) for the total charge in the
ring geometry the last term vanishes in the limits $\alpha _{0}\rightarrow
\pm 1/2$ (the real and imaginary parts in Eq. (\ref{j02bn}) vanish
separately) and, hence, similarly to the single boundary part, the total
charge density $\langle j^{0}\rangle $ is a continuous function at the
half-odd integer values of $\alpha $.

\begin{figure}[tbph]
\begin{center}
\begin{tabular}{cc}
\epsfig{figure=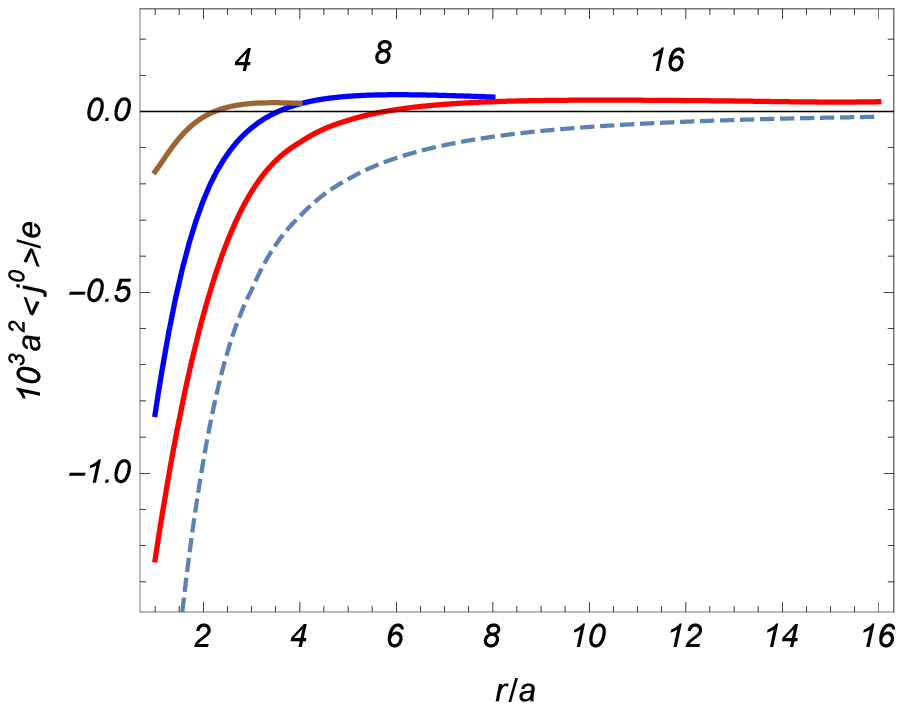,width=7.cm,height=5.5cm} & \quad %
\epsfig{figure=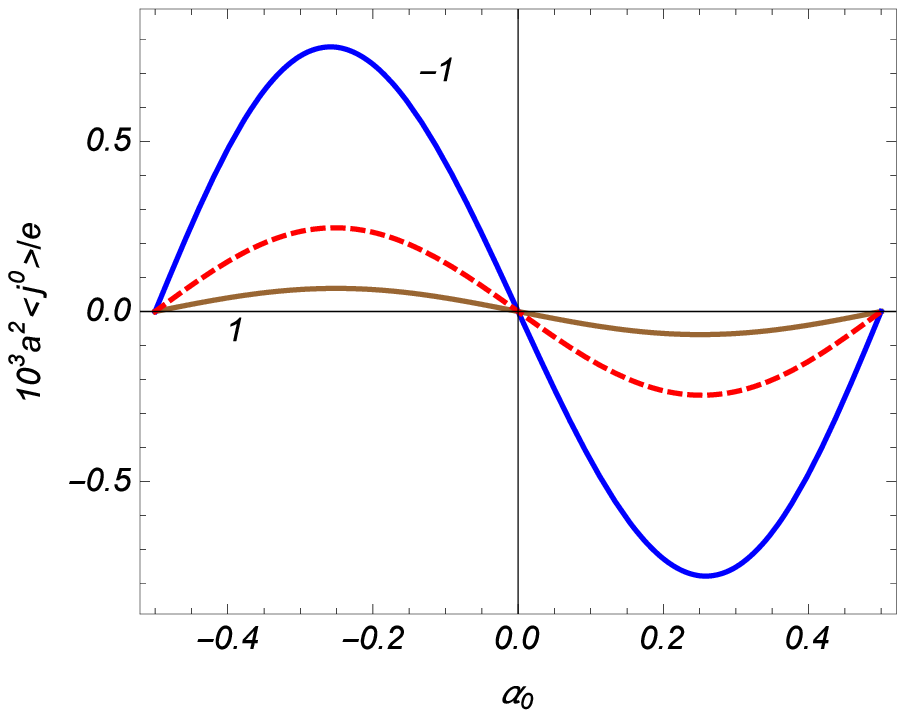,width=7.cm,height=5.5cm}%
\end{tabular}%
\end{center}
\caption{Charge density in the ring as a function of the radial coordinate
for a massless field (left panel) and as a function of the parameter $%
\protect\alpha _{0}$ (right panel). The left panel is plotted for the
magnetic flux parameter $\protect\alpha _{0}=1/4$ and the numbers near the
curves are the values of $b/a$. The dashed curve in that panel corresponds
to the charge density outside a single boundary at $r=a$. The full curves in
the right panel are plotted for $ma=0.1$, $b/a=8$, $r/a=2$ and the numbers
near the curves are the values of $s$. The dashed curve in the right panel
corresponds to a massless field.}
\label{fig2}
\end{figure}

Having discussed general features of the charge density, for the further
clarification of the dependence on the parameters of the model, let us
consider numerical examples. In the left panel of figure \ref{fig2}, for the
geometry of two boundaries at $r=a,b$, we have plotted the charge density as
a function of the radial coordinate for a massless fermionic field and for $%
\alpha _{0}=1/4$. The numbers near the curves are the values of the ratio $%
b/a$. The dashed curve presents the charge density in the geometry of a
single boundary at $r=a$, namely, the quantity $10^{3}a^{2}\langle
j^{0}\rangle _{a}^{\mathrm{(b)}}/e$. In the right panel of figure \ref{fig2}%
, by the full curves, the charge density is plotted as a function of $\alpha
_{0}$ for fixed values $ma=0.1$, $b/a=8$, $r/a=2$. The numbers near the
curves are the values of the parameter $s$. The dashed curve is the
corresponding charge density for a massless field with the same values of
the other parameters. As is seen from the left panel, the charge density is
peaked around the inner edge of the ring. The ratio $\langle j^{0}\rangle /e$
is negative near the inner edge and positive near the outer edge. In the
geometry of a single boundary at $r=a$ this ratio is negative for $\alpha
_{0}>0$. We have already mentioned that the charge density vanishes for $%
|\alpha _{0}|=1/2$, a feature seen from figure \ref{fig2}.

\begin{figure}[tbph]
\begin{center}
\begin{tabular}{cc}
\epsfig{figure=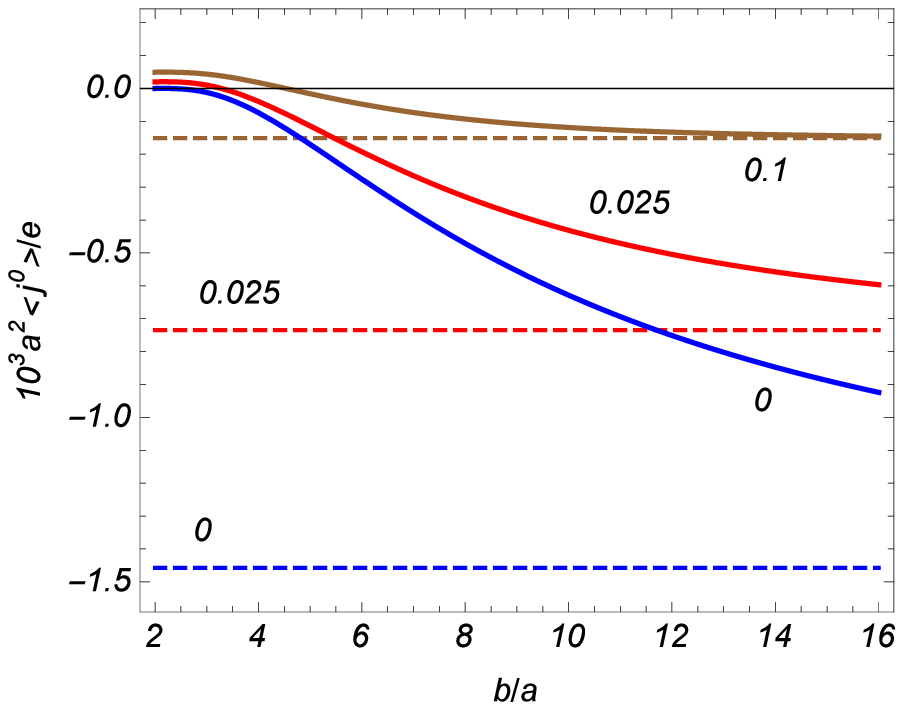,width=7.cm,height=5.5cm} & \quad %
\epsfig{figure=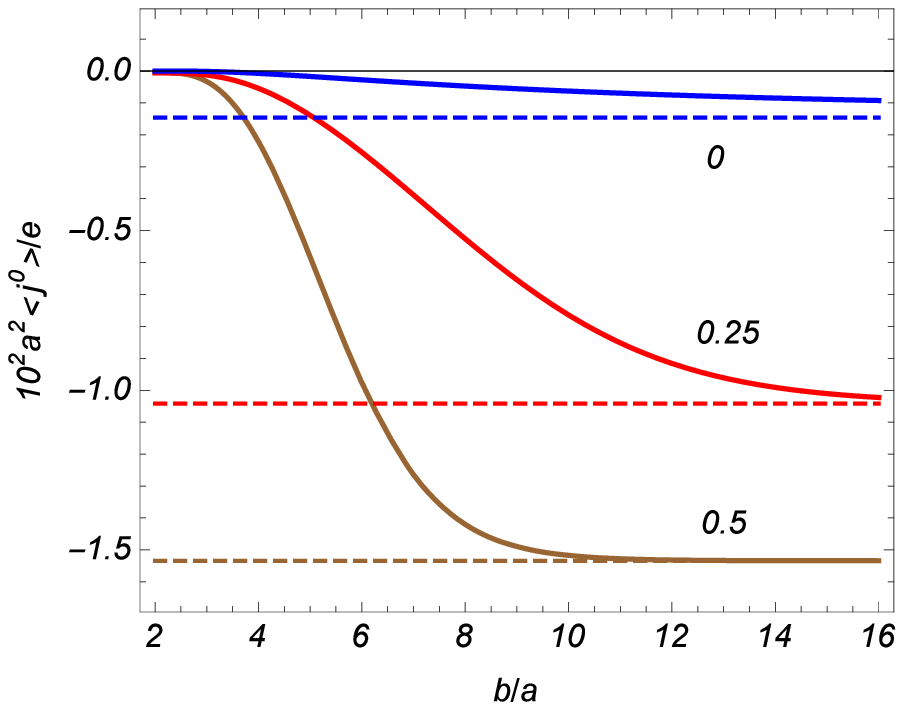,width=7.cm,height=5.5cm}%
\end{tabular}%
\end{center}
\caption{The dependence of the charge density on the ratio $b/a$ for
irreducible representations $s=1$ (left panel) and $s=-1$ (right panel). The
graphs are plotted for $\protect\alpha _{0}=1/4$, $r/a=1.5$ and the numbers
near the curves are the values for $ma$. The horizontal dashed curves
correspond to the charge density in the geometry of a single boundary at $%
r=a $.}
\label{fig3}
\end{figure}

In figure \ref{fig3} the charge density is displayed versus the ratio $b/a$
for fixed values $\alpha _{0}=1/4$, $r/a=1.5$. The numbers near the curves
are the values of the parameter $ma$. The dashed lines correspond to the
current density outside a single boundary of radius $a$. For the left and
right panels $s=1$ and $s=-1$, respectively. Note that the scaling factors
for these panels are different. The charge density for the representation $%
s=-1$ is essentially larger. The general feature seen from figure \ref{fig3}
is that the presence of the outer edge leads to the decrease of the absolute
value of the charge density. For large values of $b/a$, the approach of the
charge density in the ring geometry to the corresponding quantity in the
geometry with a single edge is quicker with increasing mass.

\begin{figure}[tbph]
\begin{center}
\begin{tabular}{cc}
\epsfig{figure=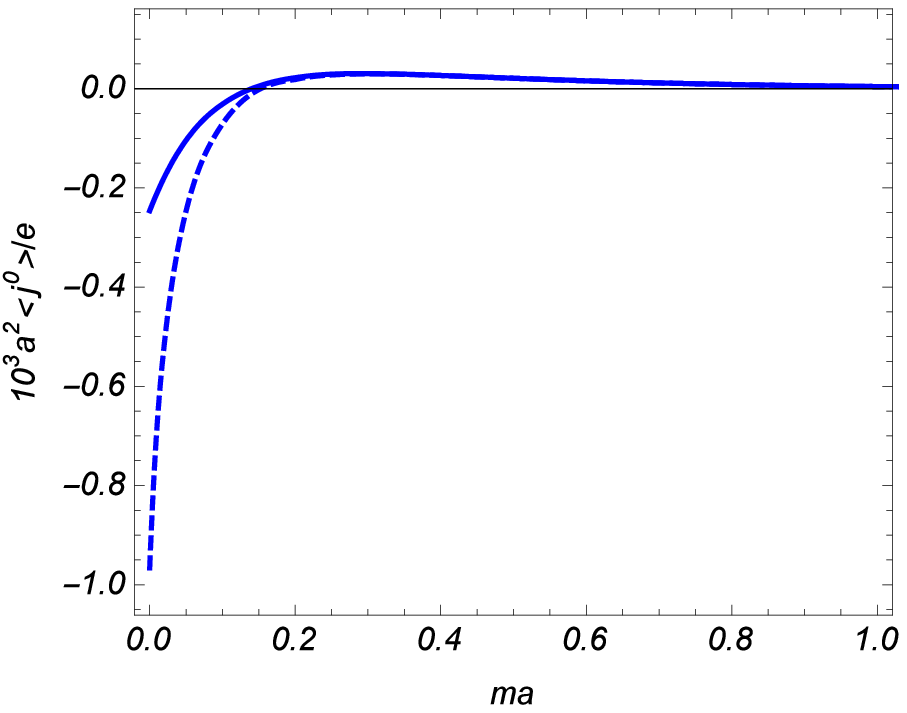,width=7.cm,height=5.5cm} & \quad %
\epsfig{figure=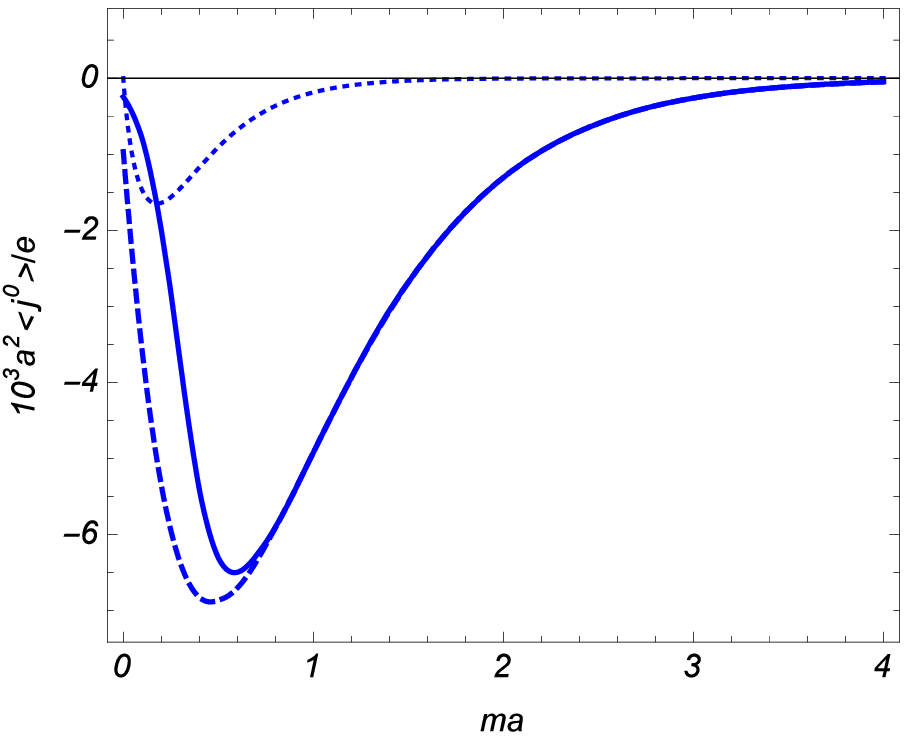,width=7.cm,height=5.5cm}%
\end{tabular}%
\end{center}
\caption{Charge density as a function of the field mass (in units of $1/a$)
for fields with $s=1$ (left panel) and $s=-1$ (right panel). For the values
of the parameters we have taken $\protect\alpha _{0}=1/4$, $b/a=8$, $r/a=2$.
The dashed curves present the charge density outside a single boundary at $%
r=a$. The dotted line corresponds to the charge density in the boundary-free
geometry.}
\label{fig4}
\end{figure}

From figures \ref{fig2} and \ref{fig3} we see that the behavior of the
charge density when the parameter $ma$ increases from $ma=0$ is essentially
different for the representations $s=1$ and $s=-1$. With the initial
increase of $ma$, the modulus of the charge density decreases for the former
case and increases for the latter one. Of course, we expect that for $ma\gg
1 $ the charge density will be suppressed for both the cases. This is seen
from figure \ref{fig4}. It presents the dependence of the charge density on
the mass of the field for irreducible representations $s=1$ (left panel) and
$s=-1$ (right panel). The graphs are plotted for $\alpha _{0}=1/4$, $b/a=8$,
$r/a=2$. The dashed curves correspond to the charge density in the geometry
of a single boundary at $r=a$. The dotted line in the right panel is the
charge density for $s=-1$ in the boundary-free problem. The corresponding
charge density for $s=1$ differs by the sign. The suppression of the VEV
with increasing $ma$ is stronger in the case $s=1$.

The representation (\ref{j02bn}) for the charge density in the ring is not
symmetric with respect to the inner and outer edges. An alternative
representation, with the extracted outer boundary part is obtained from Eq. (%
\ref{j02}) by making use of the relation%
\begin{equation}
\frac{I_{n_{p}}^{(a)}(ax)}{K_{n_{p}}^{(a)}(ax)}K_{\mu }^{2}(rx)=\frac{%
K_{n_{p}}^{(b)}(bx)}{I_{n_{p}}^{(b)}(bx)}I_{\mu
}^{2}(rx)+\sum_{u=a,b}n_{u}\Omega _{un_{p}}(ax,bx)G_{n_{p},\mu
}^{(u)2}(ux,rx),  \label{Rel3}
\end{equation}%
with the notation%
\begin{equation}
\Omega _{bn_{p}}(ax,bx)=\frac{I_{n_{p}}^{(a)}(ax)/I_{n_{p}}^{(b)}(bx)}{%
K_{n_{p}}^{(a)}\left( ax\right) I_{n_{p}}^{(b)}\left( bx\right)
-I_{n_{p}}^{(a)}(ax)K_{n_{p}}^{(b)}(bx)}.  \label{Omb}
\end{equation}%
The expression for the charge density takes the form%
\begin{eqnarray}
\langle j^{0}\rangle &=&\langle j^{0}\rangle _{b}+\frac{e}{2\pi ^{2}}%
\sum_{n=0}^{\infty }\sum_{p=\pm }p\,\int_{m}^{\infty }dx\,x\left\{ \frac{sm}{%
\sqrt{x^{2}-m^{2}}}\right.  \notag \\
&&\times \mathrm{Re}\left[ \Omega _{bn_{p}}(ax,bx)\left(
G_{n_{p},n_{p}}^{(b)2}(bx,rx)+G_{n_{p},n_{p}+1}^{(b)2}(bx,rx)\right) \right]
\notag \\
&&\left. +\mathrm{Im}\left[ \Omega _{bn_{p}}(ax,bx)\left(
G_{n_{p},n_{p}+1}^{(b)2}(bx,rx)-G_{n_{p},n_{p}}^{(b)2}(bx,rx)\right) \right]
\right\} .  \label{j02bnp}
\end{eqnarray}%
Here, the first term in the right-hand side is decomposed as%
\begin{equation}
\langle j^{0}\rangle _{b}=\langle j^{0}\rangle _{0}+\langle j^{0}\rangle
_{b}^{\mathrm{(b)}},  \label{j0b}
\end{equation}%
with%
\begin{eqnarray}
\langle j^{0}\rangle _{b}^{\mathrm{(b)}} &=&\frac{e}{2\pi ^{2}}%
\sum_{n=0}^{\infty }\sum_{p=\pm }p\,\,\int_{m}^{\infty }dx\,x\left\{ sm\frac{%
I_{n_{p}}^{2}(rx)+I_{n_{p}+1}^{2}(rx)}{\sqrt{x^{2}-m^{2}}}\mathrm{Re}\left[
\frac{K_{n_{p}}^{(b)}(bx)}{I_{n_{p}}^{(b)}(bx)}\right] \right.  \notag \\
&&\left. +\left[ I_{n_{p}+1}^{2}(rx)-I_{n_{p}}^{2}(rx)\right] \mathrm{Im}%
\left[ \frac{K_{n_{p}}^{(b)}(bx)}{I_{n_{p}}^{(b)}(bx)}\right] \right\} ,
\label{j0b2n}
\end{eqnarray}%
and with the notations defined in accordance with Eq. (\ref{fnun}). For the
ratio under the signs of the real and imaginary parts in Eq. (\ref{j0b2n})
we have the following explicit expression%
\begin{equation}
\frac{K_{n_{p}}^{(u)}(z)}{I_{n_{p}}^{(u)}(z)}=\frac{W_{n_{p}}^{(u)}(z)+in_{u}%
\sqrt{1-m_{u}^{2}/z^{2}}}{z[I_{n_{p}+1}^{2}\left( z\right)
+I_{n_{p}}^{2}\left( z\right) ]+2n_{u}sm_{u}I_{n_{p}}\left( z\right)
I_{n_{p}+1}\left( z\right) }.  \label{KI}
\end{equation}%
The denominator in this expression is positive for $z\geqslant m_{u}$.
Relatively simple expressions for single boundary parts (\ref{j0np}) and (%
\ref{j0b2n}) are obtained for a massless field.

If $|\alpha _{0}|\neq 1/2$, in the limit $a\rightarrow 0$ one has%
\begin{equation}
\Omega _{bn_{p}}(az,bz)\approx \frac{I_{n_{p}}^{(a)}(az)}{%
K_{n_{p}}^{(a)}\left( az\right) I_{n_{p}}^{(b)2}\left( bz\right) },
\label{Omba0}
\end{equation}%
with the ratio of the modified Bessel functions given by Eq. (\ref{IKrata0}%
). From here it follows that in this limit the last term in Eq. (\ref{j02bnp}%
) vanishes. This means that the part (\ref{j0b}) is the charge density in
the region $0\leqslant r\leqslant b$ for the geometry of a single boundary
at $r=b$. The contribution (\ref{j0b2n}) is induced by the latter. Note
that, for $|\alpha _{0}|\neq 1/2$, this contribution is also obtained from
Eq. (\ref{j02bn}) in the limit $a\rightarrow 0$. Hence, we conclude that the
last term in Eq. (\ref{j02bnp}) is the contribution in the charge density
induced by adding the boundary at $r=a$ in the geometry with a single
boundary at $r=b$ (the second boundary-induced part). In the limit $%
r\rightarrow 0$ the dominant contribution in Eq. (\ref{j0b2n}) for the
single boundary contribution comes from the term with $n=0$, $p=-\mathrm{sgn}%
(\alpha _{0})$, and the boundary-induced charge density behaves as $%
1/r^{2|\alpha _{0}|}$. For a massive field the boundary-free contribution
diverges like $1/r$ and it dominates in the total VEV for $|\alpha _{0}|<1/2$%
. All the separate terms in the representation (\ref{j02bnp}) are
discontinuous at half-odd integer values of $\alpha $. However, as it has
been already emphasized before, the total charge density is a continuous
function of the magnetic flux everywhere.

\begin{figure}[tbph]
\begin{center}
\begin{tabular}{cc}
\epsfig{figure=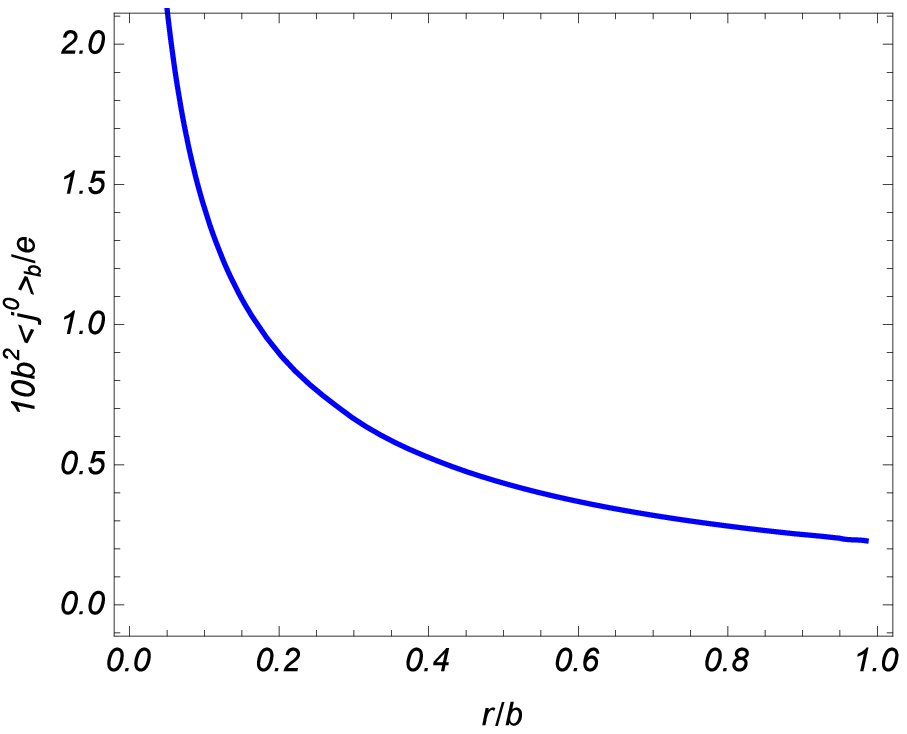,width=7.cm,height=5.5cm} & \quad %
\epsfig{figure=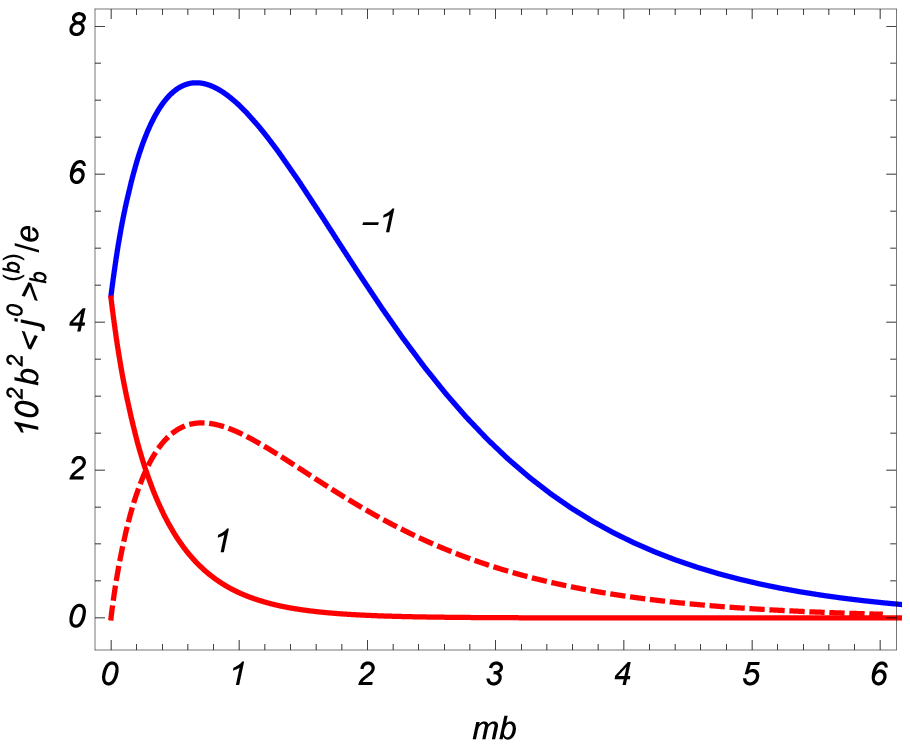,width=7.cm,height=5.5cm}%
\end{tabular}%
\end{center}
\caption{The boundary-induced part in the charge density inside a single
boundary with radius $b$ as a function of the radial coordinate (left panel)
and of the field mass (right panel) for $\protect\alpha _{0}=1/4$. The left
panel is plotted for a massless field. For the right panel $r/b=0.5$ and the
numbers near the curves are the values for $s$. The dashed curve presents
the charge density in the boundary-free geometry for $s=1$. }
\label{fig5}
\end{figure}

Figure \ref{fig5} presents the charge density in the region $r\leqslant b$
for the geometry with a single edge at $r=b$. For the corresponding magnetic
flux we have taken $\alpha _{0}=1/4$. In the left panel the charge density
is plotted versus the radial coordinate for a massless field. The right
panel displays the boundary-induced part in the charge density as a function
of the field mass in the cases $s=1$ and $s=-1$ (numbers near the curves)
and for $r/a=0.5$. The dashed curve is for the charge density in the
boundary-free geometry for $s=1$.

As we have mentioned, the last terms in the representations (\ref{j02bn})
and (\ref{j02bnp}) are induced when we add the second edge to the geometry
with a single boundary. These second boundary-induced contributions are
further simplified on the edges $r=a$ and $r=b$ respectively. The
corresponding expressions can be written in a combined form%
\begin{equation}
\langle j^{0}\rangle =\langle j^{0}\rangle _{u}+\frac{e}{\pi ^{2}}%
\sum_{n=0}^{\infty }\,\sum_{p=\pm }p\int_{m}^{\infty }dx\,\frac{x}{\sqrt{%
x^{2}-m^{2}}}\mathrm{Re}\left[ (sm+i\sqrt{x^{2}-m^{2}})\Omega
_{un_{p}}(ax,bx)\right] ,  \label{j0edge}
\end{equation}%
on the edge $r=u$ with $u=a,b$. The second term in the right-hand side
presents the charge induced on the edge $r=u$ by the other edge.

An important point to be mentioned here is that the VEV\ of the charge
density is finite on the edges (note that for the evaluation of the
corresponding limiting value of the single boundary-induced parts $\langle
j^{0}\rangle _{u}^{\mathrm{(b)}}$, $u=a,b$, we cannot directly put $r=u$ in
the representations (\ref{j0np}) and (\ref{j0b2n}) as the separate integrals
in the summation over $p$ logarithmically diverge in the upper limit). From
the theory of the Casimir effect it is known that in quantum field theory
with boundary conditions imposed on the field operator the VEVs of local
physical observables, in general, diverge on the boundaries. For example,
the latter is the case for the VEV of the energy-momentum tensor or for the
fermion condensate in problems with fermionic fields. The appearance of
surface divergences in this type of quantities is a consequence of the
idealization replacing the physical interaction by the imposition of
boundary conditions and indicates that a more realistic physical model
should be employed. For example, the microstructure of the boundary on small
scales can introduce a physical cutoff needed to produce finite values for
surface quantities.

\section{Azimuthal current}

\label{sec:Current}

Now we turn to the spatial components of the fermionic current density.
First of all, by using the mode functions (\ref{psip}) it is seen that the
mode-sum (\ref{VEVj1}) for the radial component of the current density
vanishes, $\langle j^{1}\rangle =0$, and the only nonzero component
corresponds to the azimuthal current. The corresponding mode-sum for the
physical component $\langle j_{\phi }\rangle =r\langle j^{2}\rangle $ is
presented in the form%
\begin{equation}
\langle j_{\phi }\rangle =-\frac{\pi e}{16a^{3}}\sum_{j}\,\sum_{\kappa =\pm
}\epsilon _{j}\sum_{l=1}^{\infty }T_{\beta _{j}}^{ab}(\eta
,z_{l})h_{2}(z_{l}).  \label{j2}
\end{equation}%
with the function%
\begin{equation}
h_{2}(z)=z^{2}\frac{g_{\beta _{j},\beta _{j}}(z,zr/a)}{\sqrt{%
z^{2}/a^{2}+m^{2}}}g_{\beta _{j},\beta _{j}+\epsilon _{j}}(z,zr/a).
\label{h2}
\end{equation}%
As before, the terms $\kappa =-$ and $\kappa =+$ are the contributions from
the negative- and positive-energy modes.

For the separation of the effects induced by the boundaries we apply to the
series over $l$ the summation formula (\ref{Sum}) with $h(z)=h_{2}(z)$. For
the function (\ref{h2}) one gets%
\begin{equation}
h_{2}(ze^{pi\frac{\pi }{2}})=-\frac{4}{\pi ^{2}}\frac{pi\epsilon
_{j}z^{2}G_{\beta _{j}}^{(ap)}(z,zr/a)}{\sqrt{(ze^{p\pi i/2})^{2}/a^{2}+m^{2}%
}}G_{\beta _{j}+\epsilon _{j}}^{(ap)}(z,zr/a).  \label{h2i}
\end{equation}%
By taking into account Eq. (\ref{Square}), we conclude that in the last
integral of Eq. (\ref{Sum}) for the integration range $[0,ma]$ the terms
with $p=+$ and $p=-$ cancel each other. For the integration range $%
[ma,\infty )$ of the remaining integral $\kappa $ and $p$ appear in the form
of the product $\kappa p$ and, hence, the negative- and positive-energy
modes give the same contribution to the last term. As a consequence, the
current density is presented as%
\begin{eqnarray}
\langle j_{\phi }\rangle &=&\langle j_{\phi }\rangle _{a}-\frac{e}{\pi ^{2}}%
\sum_{j}\,\int_{m}^{\infty }dx\,\frac{x^{2}}{\sqrt{x^{2}-m^{2}}}  \notag \\
&&\times \mathrm{Re}\left[ \Omega _{a\beta _{j}}(ax,bx)G_{\beta
_{j}}^{(a)}(ax,rx)G_{\beta _{j}+\epsilon _{j}}^{(a)}(ax,rx)\right] ,
\label{j22}
\end{eqnarray}%
where%
\begin{equation}
\langle j_{\phi }\rangle _{a}=-\frac{e}{4\pi a^{3}}\sum_{j}\,\sum_{\kappa
=\pm }\int_{0}^{\infty }dx\,\frac{\epsilon _{j}h_{2}(x)}{J_{\beta
_{j}}^{(a)2}(x)+Y_{\beta _{j}}^{(a)2}(x)}.  \label{j2a}
\end{equation}%
The part (\ref{j2a}) comes from the first term in the right-hand side of the
summation formula (\ref{Sum}). By taking into account that in the limit $%
b\rightarrow \infty $ the last term in Eq. (\ref{j22}) vanishes, we conclude
that $\langle j_{\phi }\rangle _{a}$ is the current density in the region $%
r\geqslant a$ for the geometry with a single boundary at $r=a$.

For the separation of the boundary-induced effects in Eq. (\ref{j2a}) we use
the relation%
\begin{equation}
\frac{g_{\beta _{j},\beta _{j}}(z,y)g_{\beta _{j},\beta _{j}+\epsilon
_{j}}(z,y)}{J_{\beta _{j}}^{(a)2}(x)+Y_{\beta _{j}}^{(a)2}(x)}=J_{\beta
_{j}}(y)J_{\beta _{j}+\epsilon _{j}}(y)-\frac{1}{2}\sum_{l=1,2}\frac{%
J_{\beta _{j}}^{(a)}(z)}{H_{\beta _{j}}^{(al)}(z)}H_{\beta
_{j}}^{(l)}(y)H_{\beta _{j}+\epsilon _{j}}^{(l)}(y).  \label{Rel2}
\end{equation}%
The further transformations are similar to that for the charge density.
Substituting Eq. (\ref{Rel2}) into Eq. (\ref{j2a}), in the part with the
last term we rotate the integration contour by the angle $\pi /2$ for the
term with $l=1$ and by the angle $-\pi /2$ for the $l=2$ term. The integrals
over the intervals $[0,ima]$ and $[0,-ima]$ are canceled. As a result, the
contribution (\ref{j2a}) is presented in the decomposed form%
\begin{equation}
\langle j_{\phi }\rangle _{a}=\langle j_{\phi }\rangle _{0}+\langle j_{\phi
}\rangle _{a}^{\mathrm{(b)}},  \label{j2a1}
\end{equation}%
where the separate terms are given by the expressions%
\begin{equation}
\langle j_{\phi }\rangle _{0}=-\frac{e}{2\pi }\sum_{j}\,\epsilon
_{j}\int_{0}^{\infty }dx\,x^{2}\frac{J_{\beta _{j}}(rx)J_{\beta
_{j}+\epsilon _{j}}(rx)}{\sqrt{x^{2}+m^{2}}},  \label{j20}
\end{equation}%
and
\begin{equation}
\langle j_{\phi }\rangle _{a}^{\mathrm{(b)}}=\frac{e}{\pi ^{2}}%
\sum_{j}\,\int_{m}^{\infty }dx\,\frac{x^{2}}{\sqrt{x^{2}-m^{2}}}\mathrm{Re}%
\left[ \frac{I_{\beta _{j}}^{(a)}(ax)}{K_{\beta _{j}}^{(a)}(ax)}\right]
K_{\beta _{j}}(rx)K_{\beta _{j}+\epsilon _{j}}(rx).  \label{j2ab}
\end{equation}%
For $\alpha $ different from a half-odd integer, the part $\langle j_{\phi
}\rangle _{a}^{\mathrm{(b)}}$ vanishes in the limit $a\rightarrow 0$ and,
hence, in this case Eq. (\ref{j20}) is interpreted as the current density in
two dimensional space without boundaries. Respectively, the part (\ref{j2ab}%
) presents the contribution induced in the region $r\leqslant a$ by a single
boundary at $r=a$. In the special case $s=1$, the single boundary-induced
contribution coincides with the result previously obtained in \cite{Beze10}
for a more general geometry of a conical space (with the sign difference
explained above).

The boundary-free part in the current density, given by Eq. (\ref{j20}),
does not depend on the parameter $s$ and, hence, on the irreducible
representation of the Clifford algebra in (2+1)-dimensional spacetime. A
more convenient expression for this part is provided in Ref. \cite{Beze10}%
\begin{equation}
\langle j_{\phi }\rangle _{0}=\frac{e\sin \left( \pi \alpha _{0}\right) }{%
4\pi ^{2}r^{2}}\int_{0}^{\infty }dz\,\frac{\cosh (2\alpha _{0}z)}{\cosh ^{3}z%
}\left( 1+2mr\cosh z\right) e^{-2mr\cosh z}.  \label{j20b}
\end{equation}%
An alternative expression is given in \cite{Flek91}:%
\begin{equation}
\langle j_{\phi }\rangle _{0}=-\frac{er}{\pi ^{3}}\sin (\pi \alpha
_{0})\int_{m}^{\infty }dx\,x^{3}\frac{K_{\alpha _{0}}^{2}(rx)-K_{1-\alpha
_{0}}(rx)K_{1+\alpha _{0}}(rx)}{\sqrt{x^{2}-m^{2}}}.  \label{j20c}
\end{equation}%
Unlike the case of the charge density, the current density does not vanish
for a massless field. For a massive field, at distances $mr\gg 1$, it
behaves as $e^{-2mr}/r^{3/2}$. At the origin the current density diverges
like $1/r^{2}$. Similarly to the case of the charge density in the
boundary-free geometry, $\langle j_{\phi }\rangle _{0}$ is discontinuous at
half-odd integer values of $\alpha $ with the discontinuity $2\langle
j_{\phi }\rangle _{0}|_{\alpha _{0}=1/2}$. In particular, for a massless
field for the discontinuity one has $e/(2\pi ^{2}r^{2})$.

Decomposing the parameter $\alpha $ in accordance with Eq. (\ref{alfdec})
and redefining the summation variable $j$, the current density is presented
in the form%
\begin{eqnarray}
\langle j_{\phi }\rangle &=&\langle j_{\phi }\rangle _{0}+\langle j_{\phi
}\rangle _{a}^{\mathrm{(b)}}-\frac{e}{\pi ^{2}}\sum_{n=0}^{\infty
}\,\sum_{p=\pm }p\,\int_{m}^{\infty }dx\,\frac{x^{2}}{\sqrt{x^{2}-m^{2}}}
\notag \\
&&\times \mathrm{Re}\left[ \Omega
_{an_{p}}(ax,bx)G_{n_{p}}^{(a)}(ax,rx)G_{n_{p}+1}^{(a)}(ax,rx)\right] ,
\label{j22n}
\end{eqnarray}%
with the single boundary-induced part%
\begin{equation}
\langle j_{\phi }\rangle _{a}^{\mathrm{(b)}}=\frac{e}{\pi ^{2}}%
\sum_{n=0}^{\infty }\,\sum_{p=\pm }p\int_{m}^{\infty }dx\,\frac{x^{2}}{\sqrt{%
x^{2}-m^{2}}}\mathrm{Re}\left[ \frac{I_{n_{p}}^{(a)}(ax)}{K_{n_{p}}^{(a)}(ax)%
}\right] K_{n_{p}}(rx)K_{n_{p}+1}(rx).  \label{j2abn}
\end{equation}%
This representation explicitly shows that the current density does not
depend on the integer part $N$ and is an odd function of the fractional part
$\alpha _{0}$. For a massless field and at large distances from the
boundary, $r\gg a$, the single boundary-induced contribution (\ref{j2abn})
behaves as $\left( a/r\right) ^{4-4|\alpha _{0}|}$, $|\alpha _{0}|<1/2$,
with the sign $\mathrm{sgn}(\alpha _{0})\langle j_{\phi }\rangle _{a}^{%
\mathrm{(b)}}/e<0$. In this limit, the total VEV\ in the geometry of a
single boundary is dominated by the boundary-free part. In the case of a
massive field, at distances $r\gg a,m^{-1}$, for the part (\ref{j2abn}) one
has the suppression by the factor $e^{-2mr}/r^{3/2}$ and the
boundary-induced contribution is of the same order as the boundary-free one.

Both the terms in the right-hand side of Eq. (\ref{j2a1}) are discontinuous
at half-odd integer values of the parameter $\alpha $. For the corresponding
limiting values one has%
\begin{equation}
\lim_{\alpha _{0}\rightarrow \pm 1/2}\langle j_{\phi }\rangle
_{0}=-\lim_{\alpha _{0}\rightarrow \pm 1/2}\langle j_{\phi }\rangle _{a}^{%
\mathrm{(b)}}=\pm \frac{em}{2\pi ^{2}r}K_{1}(2mr).  \label{j2discont}
\end{equation}%
Though the separate terms are discontinuous, the total current density in
the region $r>a$ for the geometry of a single boundary vanishes in the
limits $\alpha _{0}\rightarrow \pm 1/2$ and it is continuous. This is the
case for the current density (\ref{j22n}) in the geometry of the ring as
well. It can be checked that the last term in the right-hand side of this
formula vanishes for $\alpha _{0}\rightarrow \pm 1/2$. Hence, we conclude
that the current density in the ring is a continuous function of the
magnetic flux including the points corresponding to the half-odd integer
values of the magnetic flux in units of the flux quantum.

\begin{figure}[tbph]
\begin{center}
\begin{tabular}{cc}
\epsfig{figure=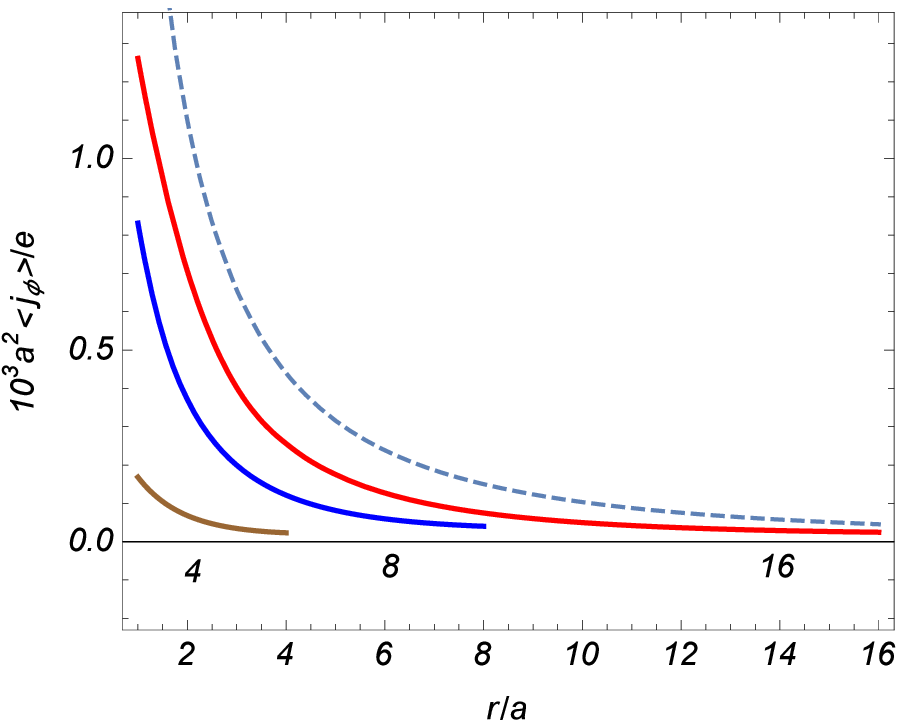,width=7.cm,height=5.5cm} & \quad %
\epsfig{figure=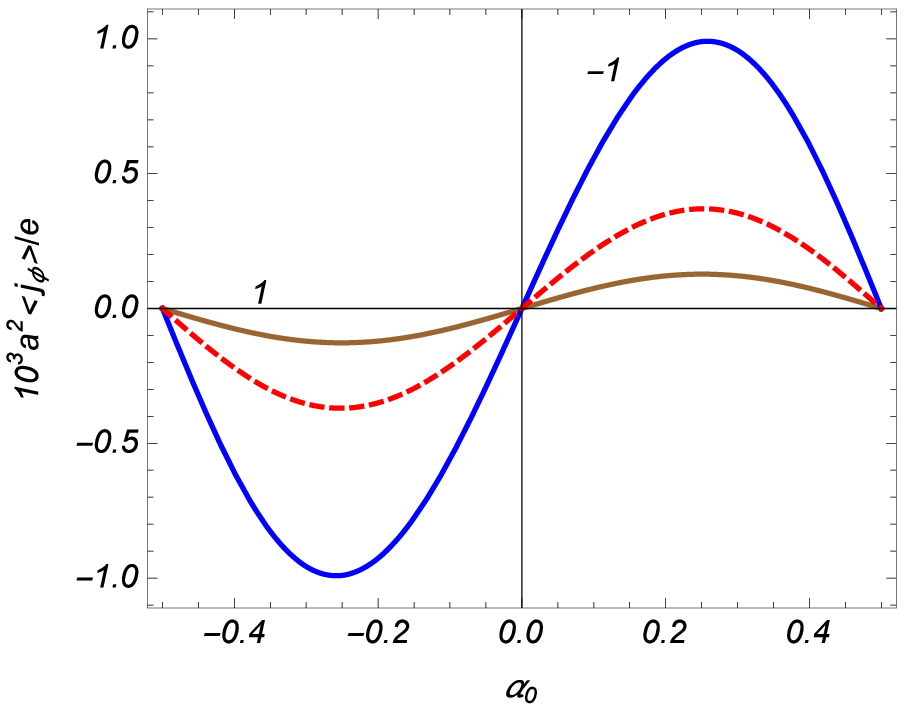,width=7.cm,height=5.5cm}%
\end{tabular}%
\end{center}
\caption{Current density in the region between two boundaries as a function
of the radial coordinate for a massless field (left panel) and as a function
of the parameter $\protect\alpha _{0}$ (right panel). The left panel is
plotted for the magnetic flux parameter $\protect\alpha _{0}=1/4$ and the
numbers near the curves are the values of $b/a$. The full curves in the
right panel are plotted for $ma=0.1$, $b/a=8$, $r/a=2$ and the numbers near
the curves are the values of $s$. The dashed curve in the right panel
corresponds to a massless field.}
\label{fig6}
\end{figure}

In order to clarify the dependence of the current density on the parameters
of the problem, let us consider numerical examples. The behavior of the
current density in the region $a\leqslant r\leqslant b$ as a function of the
radial coordinate and of the parameter $\alpha _{0}$ is presented in figure %
\ref{fig6}. The left panel is plotted for a massless field and for the
magnetic flux parameter $\alpha _{0}=1/4$. In this panel, the numbers near
the full curves are the values of the ratio $b/a$ and the dashed curve
presents the current density in the geometry of a single boundary at $r=a$.
The full curves in the right panel are plotted for $b/a=8$, $r/a=2$, $ma=0.1$
and the numbers near them are the values of the parameter $s$. The dashed
curve in the right panel corresponds to the current density for a massless
field for the same values of $b/a$ and $r/a$. Similarly to the VEV\ of the
charge density, the current density is finite on the edges. As it has been
already emphasized, the current density vanishes for half-odd integer values
of the ratio of the magnetic flux to the flux quantum, corresponding to $%
|\alpha _{0}|=1/2$. The graphs show that the current density is peaked near
the inner edge and it decreases with decreasing the width of the ring.

\begin{figure}[tbph]
\begin{center}
\begin{tabular}{cc}
\epsfig{figure=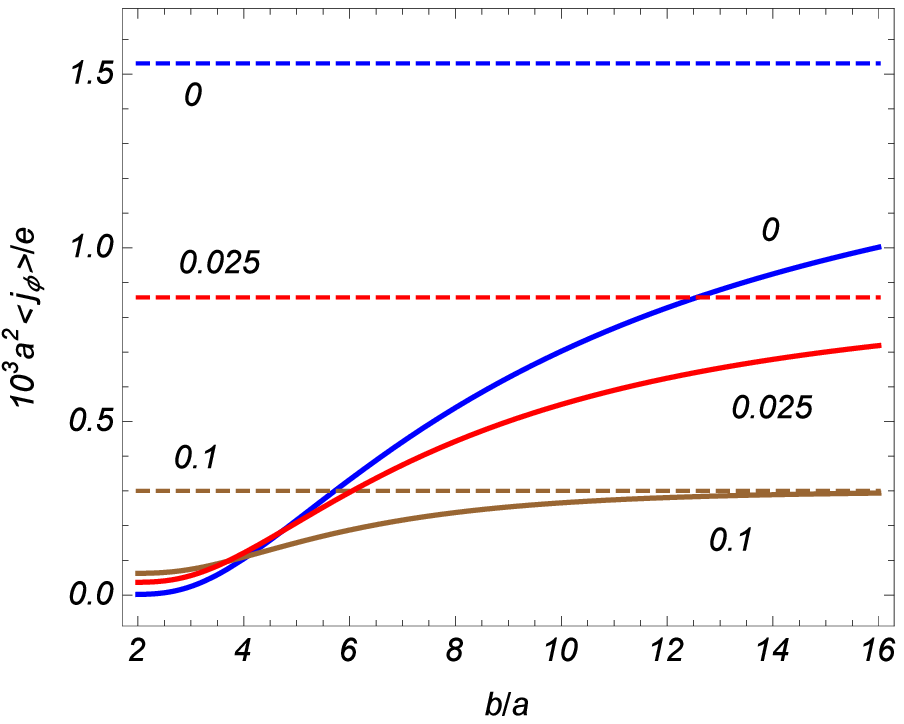,width=7.cm,height=5.5cm} & \quad %
\epsfig{figure=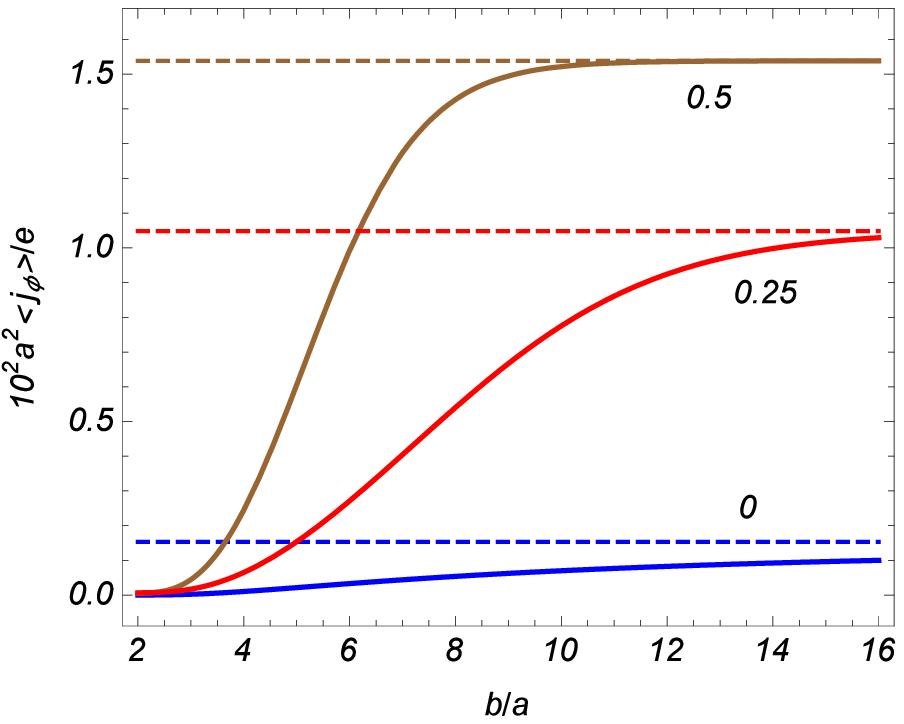,width=7.cm,height=5.5cm}%
\end{tabular}%
\end{center}
\caption{Current density versus the ratio $b/a$ for irreducible
representations $s=1$ (left panel) and $s=-1$ (right panel). The graphs are
plotted for $\protect\alpha _{0}=1/4$, $r/a=1.5$ and the numbers near the
curves are the values of $ma$. The dashed curves present the current density
in the geometry of a single boundary at $r=a$.}
\label{fig7}
\end{figure}

In figure \ref{fig7}, the current density is displayed as a function of the
relative location of the outer boundary for fixed values of the parameters $%
\alpha _{0}=1/4$, $r/a=1.5$. The numbers near the curves are the values for $%
ma$. The left and right panels correspond to the representations with $s=1$
and $s=-1$, respectively. The dashed lines on both the panels present the
current density in the geometry of a single boundary at $r=a$. Again, the
graphs show that, for a fixed inner radius, the current density increases
with increasing the width of the ring.

\begin{figure}[tbph]
\begin{center}
\begin{tabular}{cc}
\epsfig{figure=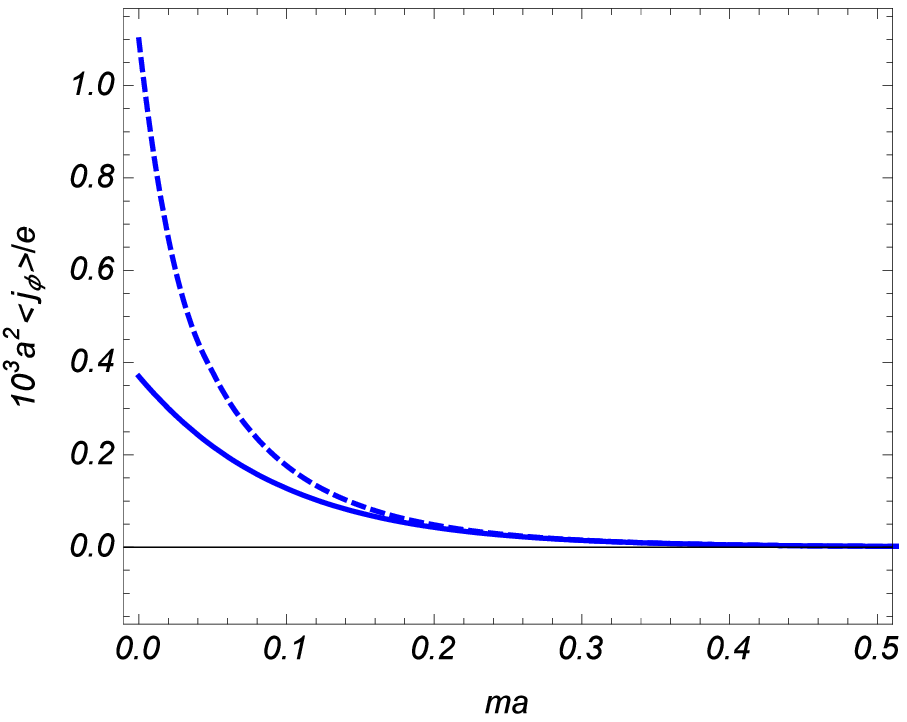,width=7.cm,height=5.5cm} & \quad %
\epsfig{figure=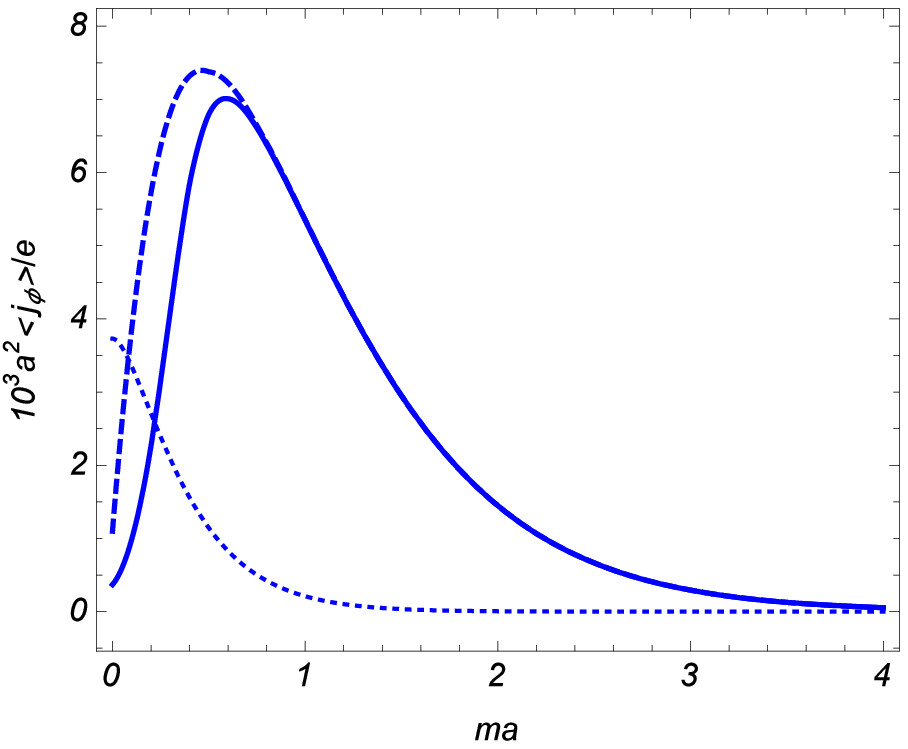,width=7.cm,height=5.5cm}%
\end{tabular}%
\end{center}
\caption{Current density versus the field mass for the fields with $s=1$
(left panel) and $s=-1$ (right panel) and for $\protect\alpha _{0}=1/4$, $%
b/a=8$, $r/a=2$. The dashed curves present the current density outside a
single boundary at $r=a$ and the dotted line is the current density in the
boundary-free geometry.}
\label{fig8}
\end{figure}

Similarly to the case of the charge density, we see an essentially different
behavior of the current density for the representations $s=1$ and $s=-1$, as
a function of the field mass. The dependence of the current density on the
mass is plotted in figure \ref{fig8} for the irreducible representations $s=1
$ (left panel) and $s=-1$ (right panel) and for the values of the parameters
$\alpha _{0}=1/4$, $b/a=8$, $r/a=2$. The dashed curves present the charge
density in the geometry of a single boundary at $r=a$. The dotted line in
the right panel is the current density in the boundary-free problem.

In the representation (\ref{j22n}) the current density for the geometry of a
single boundary at $r=a$ is explicitly separated. The representation with
the outer edge contribution separated is obtained by using the identity%
\begin{eqnarray}
&&-\frac{I_{n_{p}}^{(a)}(ax)}{K_{n_{p}}^{(a)}(ax)}%
K_{n_{p}}(rx)K_{n_{p}+1}(rx)=\frac{K_{n_{p}}^{(b)}(bx)}{I_{n_{p}}^{(b)}%
\left( bx\right) }I_{n_{p}}\left( rx\right) I_{n_{p}+1}\left( rx\right)
\notag \\
&&\qquad +\sum_{u=a,b}n_{u}\Omega
_{un_{p}}(ax,bx)G_{n_{p},n_{p}}^{(u)}(ux,rx)G_{n_{p},n_{p}+1}^{(u)}(ux,rx).
\label{Id2}
\end{eqnarray}%
With this relation, the current density in the region $a\leqslant r\leqslant
b$ is presented as%
\begin{eqnarray}
\langle j_{\phi }\rangle &=&\langle j_{\phi }\rangle _{b}-\frac{e}{\pi ^{2}}%
\sum_{n=0}^{\infty }\,\sum_{p=\pm }p\,\int_{m}^{\infty }dx\,\frac{x^{2}}{%
\sqrt{x^{2}-m^{2}}}  \notag \\
&&\times \mathrm{Re}\left[ \Omega
_{bn_{p}}(ax,bx)G_{n_{p},n_{p}}^{(b)}(bx,rx)G_{n_{p},n_{p}+1}^{(b)}(bx,rx)%
\right] .  \label{j2b}
\end{eqnarray}%
Here%
\begin{equation}
\langle j_{\phi }\rangle _{b}=\langle j_{\phi }\rangle _{0}+\langle j_{\phi
}\rangle _{b}^{\mathrm{(b)}},  \label{jphib}
\end{equation}%
with%
\begin{equation}
\langle j_{\phi }\rangle _{b}^{\mathrm{(b)}}=-\frac{e}{\pi ^{2}}%
\sum_{n=0}^{\infty }\,\sum_{p=\pm }p\,\int_{m}^{\infty }dx\,\frac{x^{2}}{%
\sqrt{x^{2}-m^{2}}}\mathrm{Re}\left[ \frac{K_{n_{p}}^{(b)}(bx)}{%
I_{n_{p}}^{(b)}(bx)}\right] I_{n_{p}}(rx)I_{n_{p}+1}(rx).  \label{j22b}
\end{equation}%
By taking into account that for $|\alpha _{0}|\neq 1/2$ one has $\Omega
_{bn_{p}}(az,bz)\sim a^{2n+1+2\alpha _{0}}$ for $a\rightarrow 0$, from Eq. (%
\ref{j2b}) we conclude that the part $\langle j_{\phi }\rangle _{b}$ is the
current density in the region $0\leqslant r\leqslant b$ for the geometry of
a single boundary at $r=b$ and the contribution (\ref{j22b}) is induced by
the boundary. For points near the center of the disc $r\leqslant b$, the
dominant contribution to the edge-induced part (\ref{j22b}) comes from the
term $n=0$ and it behaves as $r^{1-2|\alpha _{0}|}$. By taking into account
that the boundary-free part behaves like $1/r^{2}$, we see that near the
center the current density is dominated by the boundary-free part.

\begin{figure}[tbph]
\begin{center}
\begin{tabular}{cc}
\epsfig{figure=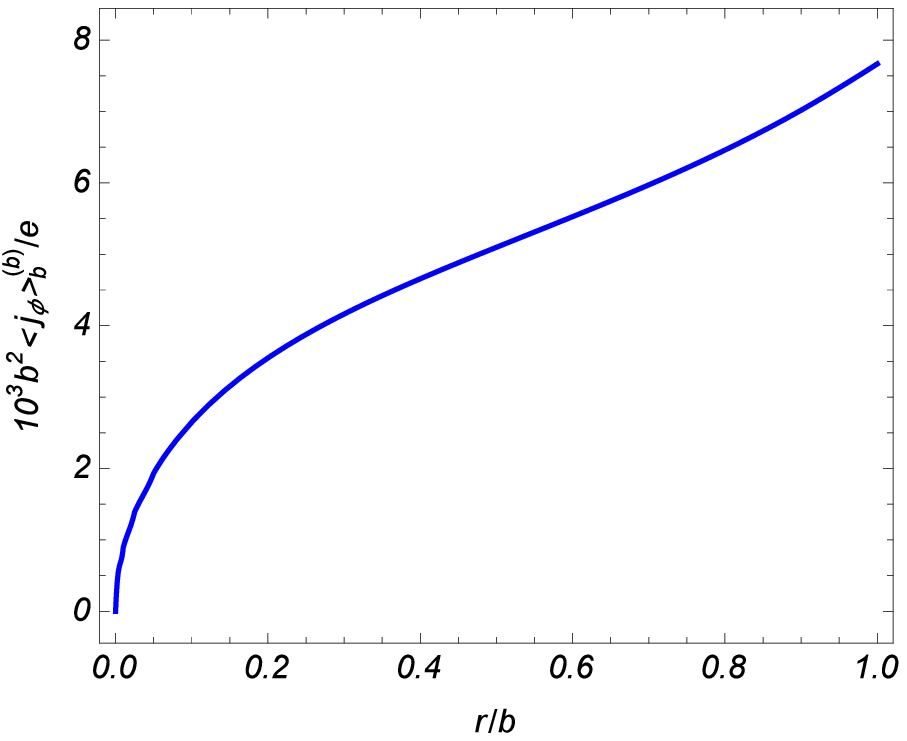,width=7.cm,height=5.5cm} & \quad %
\epsfig{figure=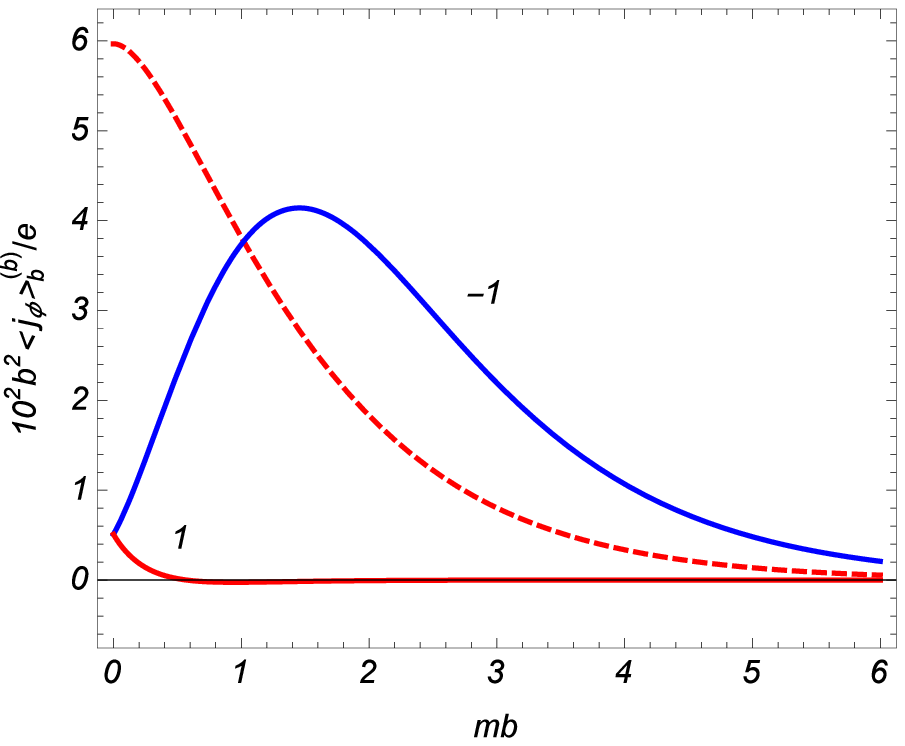,width=7.cm,height=5.5cm}%
\end{tabular}%
\end{center}
\caption{The boundary-induced contribution to the current density inside a
single circular boundary with radius $b$ versus the radial coordinate and
the mass for $\protect\alpha _{0}=1/4$. The left panel is plotted for a
massless field. For the right panel $r/b=0.5$ and the numbers near the
curves are the values of $s$.}
\label{fig9}
\end{figure}

For the magnetic flux parameter $\alpha _{0}=1/4$, the boundary-induced
charge density (\ref{j22b}) in the geometry of a single boundary at $r=b$ is
plotted in figure \ref{fig9} versus the radial coordinate and the mass. The
dashed curve corresponds to the current density in the boundary-free
problem. The left panel is plotted for a massless field and the numbers near
the curves on the right panel are the values of the parameter $s$ and we
have taken $r/b=0.5$.

The last term in Eq. (\ref{j2b}) is the current density induced in the
region $a\leqslant r\leqslant b$ when we add the boundary at $r=a$ to the
geometry of the disc with the radius $b$. All the separate terms in the
right-hand side of Eq. (\ref{j2b}) have jumps at half-odd integer values of $%
\alpha $. However, it can be seen that the total current density is a
continuous function of the magnetic flux and it vanishes for $\alpha =N+1/2$%
. Similar to the case of the charge density, relatively simple expressions
for the second edge-induced parts in the representations (\ref{j22n}) and (%
\ref{j2b}) are obtained on the edges $\ r=a$ and $r=b$, respectively. The
current density on the edge $r=u$ with $u=a,b$, is presented in the form%
\begin{equation}
\langle j_{\phi }\rangle =\langle j_{\phi }\rangle _{u}+\frac{en_{u}}{\pi
^{2}}\sum_{n=0}^{\infty }\,\sum_{p=\pm }p\,\int_{m}^{\infty }dx\,\frac{x}{%
\sqrt{x^{2}-m^{2}}}\mathrm{Re}\left[ (sm+i\sqrt{x^{2}-m^{2}})\Omega
_{un_{p}}(ax,bx)\right] .  \label{j2edge}
\end{equation}%
Comparing with Eq. (\ref{j0edge}), we see that the second edge-induced
contribution in the current density on the first edge is equal to the
corresponding charge density for the outer edge and has the opposite sign on
the inner edge. These relations between the charge and current densities on
the ring edges hold for the total VEVs as well. If we formally put $r=a,b$
in the mode sums (\ref{j0}) and (\ref{j2}), then, by using the Wronskian
relation for the Bessel and Neumann functions and Eq. (\ref{EigEq}) for the
eigenvalues of the radial quantum number, we can see that%
\begin{equation}
\langle j^{0}\rangle _{r=u}=n_{u}\langle j_{\phi }\rangle _{r=u}=-\frac{e}{%
4\pi a^{2}}\sum_{j}\sum_{l=1}^{\infty }\sum_{\kappa =\pm }\,\kappa T_{\beta
_{j}}^{ab}(\eta ,z_{l})\frac{E+\kappa sm}{E}z_{l}B_{u}(z_{l}).  \label{j0u}
\end{equation}%
with $u=a,b$ and
\begin{equation}
B_{a}(x)=1,\;B_{b}(x)=J_{\beta _{j}}^{(a)2}\left( x\right) /J_{\beta
_{j}}^{(b)2}\left( xb/a\right) .  \label{Bab}
\end{equation}%
This feature is a consequence of the bag boundary conditions we have imposed
on the edges (see also \cite{Bene12}).

In the discussion above for the charge and current densities we have assumed
that the fermionic field is in the vacuum state. If the field is in thermal
equilibrium at finite temperature, in addition to the vacuum parts we have
considered here, the expectation values will receive contributions coming
from particles and antiparticles (for finite temperature effects on the
fermionic charge and current densities in topologically nontrivial spaces
see, e.g., \cite{Bell14T}; see also \cite{Khan14} for a recent review of
finite temperature field theoretical effects in toroidal topology). For a
fermionic field with the chemical potential $\mu $, obeying the condition $%
|\mu |<m$, the charge and current densities at zero temperature coincide
with the VEVs we have investigated above. In the case $|\mu |>m$, the zero
temperature expectation values in addition to the VEVs will contain
contributions from particles or antiparticles (depending on the sign of the
chemical potential) filling the states with the energies $E$ in the range $%
m\leqslant E\leqslant |\mu |$.

\section{Charge and current densities in P- and T-symmetric models with
applications to graphene rings}

\label{sec:Tsym}

By using the results from the previous sections we can obtain the vacuum
densities in the parity and time-reversal symmetric massive fermionic
models. In (2+1) dimensions the irreducible representations for the Clifford
algebra are realized by $2\times 2$ matrices. In cylindrical coordinates, we
can choose the Dirac matrix $\gamma ^{2}$ in two inequivalent ways: $\gamma
^{2}=\gamma _{(s)}^{2}=-is\gamma ^{0}\gamma ^{1}/r$, where, as before, $%
s=\pm 1$. The gamma matrices (\ref{DirMat2}),\ we have used in the
discussion above, correspond to the representation with the upper sign. Two
sets of Dirac matrices $\gamma _{(s)}^{\mu }=(\gamma ^{0},\gamma ^{1},\gamma
_{(s)}^{2})$ realize two inequivalent irreducible representations of the
Clifford algebra. In these representations, the mass term in the Lagrangian
density for a two-component spinor field,
\begin{equation}
L_{s}=\bar{\psi}_{(s)}(i\gamma _{(s)}^{\mu }D_{\mu }-m)\psi _{(s)},
\label{Ls}
\end{equation}%
is not invariant under the $P$- and $T$-transformations. Here we assume that
both the fields $\psi _{(s)}$ obey the same boundary conditions%
\begin{equation}
\left( 1+in_{\mu }\gamma _{(s)}^{\mu }\right) \psi _{(s)}(x)=0,  \label{BCs}
\end{equation}%
on the circular edges $r=a,b$. In order to recover the $P$- and $T$%
-invariances, let us consider the combined Lagrangian density $L=\sum_{s=\pm
1}L_{s}$. By appropriate transformations of the fields $\psi _{(-1)}$ and $%
\psi _{(+1)}$, this Lagrangian is invariant under $P$- and $T$%
-transformations (in the absence of magnetic fields).

In order to relate the fields $\psi _{(s)}$ to the ones we have considered
in the evaluation of the vacuum densities, let us introduce new
two-component fields $\psi _{(s)}^{\prime }$ in accordance with $\psi
_{(+1)}^{\prime }=\psi _{(+1)}$, $\psi _{(-1)}^{\prime }=\gamma ^{0}\gamma
^{1}\psi _{(-1)}$. In terms of these fields, the Lagrangian density is
presented as%
\begin{equation}
L=\sum_{s=\pm 1}\bar{\psi}_{(s)}^{\prime }(i\gamma ^{\mu }D_{\mu }-sm)\psi
_{(s)}^{\prime },  \label{Lag2}
\end{equation}%
where $\gamma ^{\mu }=\gamma _{(+1)}^{\mu }$. From Eq. (\ref{Lag2}) we
conclude that the equations for the fields with $s=-1$ and $s=+1$ differ by
the sign of the mass term and coincide with Eq. (\ref{Direq}). From the
boundary conditions (\ref{BCs}) it follows that the fields in Eq. (\ref{Lag2}%
) obey the conditions%
\begin{equation}
\left( 1+isn_{\mu }\gamma ^{\mu }\right) \psi _{(s)}^{\prime }(x)=0,
\label{BCsn}
\end{equation}%
on $r=a,b$. As it is seen, the field $\psi _{(+1)}^{\prime }$ obeys the
condition (\ref{BCMIT}), whereas the boundary condition for the field $\psi
_{(-1)}^{\prime }$ differs by the sign of the term with the normal to the
boundary. As it has been already noticed in \cite{Berr87}, this type of
condition with the reversed sign is an equally acceptable boundary condition
for the Dirac equation. Combining $2$-component spinors $\psi _{+}^{\prime }$
and $\psi _{-}^{\prime }$ in a $4$-component one, $\Psi =(\psi _{+}^{\prime
},\psi _{-}^{\prime })^{T}$, and introducing $4\times 4$ Dirac matrices $%
\gamma _{(4)}^{\mu }=\sigma _{3}\otimes \gamma ^{\mu }$, the Lagrangian
density (\ref{Lag2}) is rewritten in the form%
\begin{equation}
L=\bar{\Psi}(i\gamma _{(4)}^{\mu }D_{\mu }-m)\Psi .  \label{Lcomb}
\end{equation}%
In this reducible representation, the boundary conditions (\ref{BCsn}) are
combined as%
\begin{equation}
\left( 1+in_{\mu }\gamma _{(4)}^{\mu }\right) \Psi (x)=0.  \label{BCcomb}
\end{equation}%
The latter has the form of the standard MIT bag condition for a 4-component
spinor.

By taking into account that $\bar{\psi}_{(s)}\gamma _{(s)}^{\mu }\psi _{(s)}=%
\bar{\psi}_{(s)}^{\prime }\gamma ^{\mu }\psi _{(s)}^{\prime }$, for the
total VEV of the current density in the model with the combined Lagrangian $%
L=\sum_{s=\pm 1}L_{s}$, with $L_{s}$ form (\ref{Ls}), one gets%
\begin{equation}
\langle J^{\mu }\rangle =\sum_{s=\pm 1}\langle \bar{\psi}_{(s)}\gamma
_{(s)}^{\mu }\psi _{(s)}\rangle =\sum_{s=\pm 1}\langle \bar{\psi}%
_{(s)}^{\prime }\gamma ^{\mu }\psi _{(s)}^{\prime }\rangle .  \label{jred}
\end{equation}%
The charge and current densities for the field $\psi _{(+1)}$ are obtained
from the expressions given in the previous sections with $s=1$. In order to
find the VEVs for the field $\psi _{(-1)}^{\prime }$, we note that it obeys
the field equation (\ref{Direq}) with $s=-1$ and the boundary condition that
differs from Eq. (\ref{BCMIT}) by the sign of the term containing the normal
to the boundaries. Consequently, the VEV $\langle \bar{\psi}_{(-1)}^{\prime
}\gamma ^{\mu }\psi _{(-1)}^{\prime }\rangle $ is obtained from the
corresponding formulas given above taking $s=-1$ and making the replacement%
\begin{equation}
n_{u}\rightarrow -n_{u},\;u=a,b  \label{nurepl}
\end{equation}%
(see Eq. (\ref{nab}) for the definition of $n_{u}$). In the final formulas
this replacement is made through the definition (\ref{fnun}) for $f=I,K$.
From here we conclude that the expressions for the VEVs $\langle \bar{\psi}%
_{(s)}^{\prime }\gamma ^{\mu }\psi _{(s)}^{\prime }\rangle $, $\mu =0,2$,
are given by the formulas in sections \ref{sec:Charge} and \ref{sec:Current}
where now we should take%
\begin{equation}
f_{n_{p}}^{(u)}(z)=\delta _{f}zf_{n_{p}+1}\left( z\right) +n_{u}(m_{u}+si%
\sqrt{z^{2}-m_{u}^{2}})f_{n_{p}}\left( z\right) ,  \label{fnun2}
\end{equation}%
with $n_{u}$ defined in accordance with Eq. (\ref{nab}). This shows that one
has the relation $f_{n_{p}}^{(u)}(z)|_{s=-1}=f_{n_{p}}^{(u)\ast }(z)|_{s=1}$
and, hence, the same for the functions $\Omega _{un_{p}}(ax,bx)$ and $%
G_{n_{p},\mu }^{(u)}(ux,rx)$. Here the star stands for the complex
conjugate. Assuming that the masses $m$ for the fields with $s=+1$ and $s=-1$
are the same, we can see that the boundary-induced contributions from these
fields to the charge density cancel each other. By taking into account that
the same is the case for the boundary-free part (see Eq. (\ref{j002c})), we
conclude that the VEV of the total charge density vanishes. For the VEV\ of
the current density, the contributions from the fields $\psi _{(+1)}^{\prime
}$ and $\psi _{(-1)}^{\prime }$ coincide for both the boundary-free and
boundary-induced parts. The corresponding expressions for the total current (%
\ref{jred}) are obtained from those in the previous section for the case $%
s=1 $ with an additional factor 2. Now, for the field $\psi _{(s)}^{\prime }$
the analog of the relation (\ref{j0u}) between the charge and current
densities on the edges has the form%
\begin{equation}
\langle j_{\phi }^{\prime }\rangle _{r=u}=-n_{u}\langle j^{0\prime }\rangle
_{r=u},  \label{j0up}
\end{equation}%
where $\langle j^{\mu \prime }\rangle =\langle \bar{\psi}_{(-1)}^{\prime
}\gamma ^{\mu }\psi _{(-1)}^{\prime }\rangle $, $u=a,b$, and $n_{u}$ is the
same as in Eq. (\ref{j0u}).

Note that we could consider another class of $P$- and $T$-invariant models
with two-components fields $\psi _{(s)}$ obeying the boundary conditions in
which the sign of the term in Eq. (\ref{BCs}), containing the normal vector,
is reversed. For these models, the same reversion should be made in the
boundary condition (\ref{BCsn}) for the primed fields. Now we see that the
charge and current densities for the field $\psi _{(-1)}^{\prime }$ coincide
with those in the previous sections for $s=-1$, whereas the results for the
field $\psi _{(+1)}^{\prime }$ are obtained from those before in the case $%
s=1$ by the replacement (\ref{nurepl}). The total charge density, combined
from the fields with $s=1$ and $s=-1$ vanishes as in the previous case, and
the current density is obtained from the expressions in section \ref%
{sec:Current} in the case $s=-1$ with the additional factor 2. As we have
seen, in this case the dependence of the VEV on the field mass is more
interesting. Of course, there is another possibility when the boundary
conditions for the fields $\psi _{(s)}$ are different for $s=1$ and $s=-1$.
For example, we could impose the condition (\ref{BCs}) with the additional
factor $s$ in the term involving the normal vector. In this case the primed
fields will obey the same boundary condition (Eq. (\ref{BCsn}) with $s=1$)
and the corresponding VEVs exactly coincide with those in sections \ref%
{sec:Charge} and \ref{sec:Current}. In this variant, there is no
cancellation and the total charge density does not vanish.

The field theoretical models we have considered can be realized by various
graphene made structures. For example, the geometry with a single boundary
at $r=a$ corresponds to a circular graphene dot if the region $r<a$ is
considered and to a single circular nanohorn (or nanopore) for the region $%
r>a$. The influence of boundaries on the electronic properties of a circular
graphene quantum dot in a magnetic field has been discussed in \cite{Schn08}%
. Comparing the analytical results obtained within the continuum model to
those obtained from the tight-binding model, the authors conclude that the
Dirac model with the infinite-mass boundary condition describes rather well
its tight-binding analog and is in good qualitative agreement with
experiments. Considering different boundary conditions in the Dirac model
for graphene devices, a similar conclusion is made in Ref. \cite{Bene12}.
The Aharonov-Bohm effect and persistent currents in graphene nanorings have
been recently investigated in \cite{Rech07,Yan10}. The effect of impurity on
persistent currents in strictly one-dimensional Dirac systems is discussed
in \cite{Stic13}.

The results obtained above can be applied for the investigation of the
ground state charge and current densities in graphene rings. Graphene is a
monolayer of carbon atoms with honeycomb lattice containing two triangular
sublattices $A$ and $B$ related by inversion symmetry. The electronic
subsystem in a graphene sheet is among the most popular realizations of the
Dirac physics in two spatial dimensions (for other planar condensed-matter
systems with the low-energy excitations described by the Dirac model see
Ref. \cite{Shar06}). For a given value of spin $S=\pm 1$, the corresponding
long wavelength excitations are described in terms of 4-component spinors $%
\Psi _{S}=(\psi _{+,AS},\psi _{+,BS},\psi _{-,AS},\psi _{-,BS})^{T}$ with
the Lagrangian density (in the standard units)
\begin{equation}
L=\sum_{S=\pm 1}\bar{\Psi}_{S}(i\hbar \gamma ^{0}\partial _{t}+i\hbar
v_{F}\gamma ^{l}D_{l}-\Delta )\Psi _{S}.  \label{LagGr}
\end{equation}%
Here, $D_{l}=(\mathbf{\nabla }-ie\mathbf{A}/\hbar c)_{l}$, $l=1,2$, is the
spatial part of the gauge extended covariant derivative and $e=-|e|$ for
electrons. The Fermi velocity $v_{F}$ plays the role of the speed of light.
It is expressed in terms of the microscopic parameters as $v_{F}=\sqrt{3}%
a_{0}\gamma _{0}/(2\hbar )\approx 7.9\times 10^{7}$ cm/s, where $%
a_{0}\approx 1.42$ \AA\ is the inter-atomic spacing of graphene honeycomb
lattice and $\gamma _{0}\approx 2.9$ eV is the transfer integral between
first-neighbor $\pi $ orbitals. The components $\psi _{\pm ,AS}$ and $\psi
_{\pm ,BS}$ of the spinor $\Psi _{S}$ give the amplitude of the electron
wave function on sublattices $A$ and $B$. The indices $+$ and $-$ of these
components correspond to inequivalent points, $\mathbf{K}_{+}$ and $\mathbf{K%
}_{-}$, at the corners of the two-dimensional Brillouin zone (see Ref. \cite%
{Gusy07}). The energy gap $\Delta $ in Eq. (\ref{LagGr}) is related to the
corresponding Dirac mass as $\Delta =mv_{F}^{2}$. It plays an important role
in many physical applications (for the mechanisms of the gap generation in
the energy spectrum of graphene see, for example, Ref. \cite{Gusy07} and
references therein). Depending of the physical mechanism for the generation,
the energy gap may take values in the range $1\,\mathrm{meV}\lesssim \Delta
\lesssim 1\,\mathrm{eV}$.

Comparing with the discussion above, we see that the values of the parameter
$s=+1$ and $s=-1$ correspond to the $\mathbf{K}_{+}$ and $\mathbf{K}_{-}$
points of the graphene Brillouin zone and the Lagrangian density (\ref{LagGr}%
) is the analog of (\ref{Lcomb}). From here we conclude that, for a given
value of the spin $S$, the expressions for the VEVs of the charge and
current densities for separate contributions coming from the points $\mathbf{%
K}_{+}$ and $\mathbf{K}_{-}$ are obtained from the formulas in previous
sections by the replacement $m\rightarrow a_{0}^{-1}\Delta /\gamma _{F}$,
where $\gamma _{F}=\hbar v_{F}/a_{0}\approx 2.51$ eV determines the energy
scale in the model. In the expressions for the current density, an
additional factor $v_{F}$ should be added, because now the operator of the
spatial components of the current density is defined as $j^{\mu }=ev_{F}\bar{%
\psi}(x)\gamma ^{\mu }\psi (x)$, $\mu =1,2$. For a given spin $S$, the
contributions from two valleys are combined in accordance with Eq. (\ref%
{jred}). In the problem at hand, the spins $S=\pm 1$ give the same
contributions to the total VEVs. As it has been mentioned before, the charge
density vanishes as a result of cancellation of the contributions from the $%
\mathbf{K}_{+}$ and $\mathbf{K}_{-}$ points. The effective charge density
may appear if the gap generation mechanism breaks the valley symmetry and
the mass gap is different for $s=+1$ and $s=-1$. Note that this will break $%
P $- and $T$-invariances of the model.

\section{Summary}

\label{sec:Conc}

In both the field theoretical and condensed matters aspects, among the most
interesting topics in quantum field theory is the investigation of the
effects induced by gauge field fluxes on the properties of the quantum
vacuum. In the present paper we have discussed the combined effects from the
magnetic flux and boundaries on the VEVs of the fermionic charge and current
densities in a two-dimensional circular ring. The examples of graphene
nanoriboons and rings have already shown that the edge effects have
important consequences on the physical properties of planar systems. In the
problem at hand, for the field operator on the ring edges we have imposed
the bag boundary conditions. The distribution of the magnetic flux inside
the inner edge can be arbitrary. The boundary separating the ring from the
region of the location for the gauge field strength is impenetrable for the
fermionic field and the effect of the gauge field is purely topological. It
depends on the total flux alone. The latter gives rise the Aharonov-Bohm
effect for physical characteristics of the ground state. The consideration
is done for both irreducible representations of the Clifford algebra in
(2+1) dimensions. In these representations the mass term in the Dirac
equation breaks the parity and time-reversal invariances. For the evaluation
of the VEVs we have employed the method based on the direct summation over a
complete set of fermionic modes in the ring. The corresponding positive- and
negative-energy wavefunctions are given by Eq. (\ref{psip}) with the radial
functions defined by Eq. (\ref{ge}). The eigenvalues of the radial quantum
number are quantized by the boundary conditions and are roots of the
equation (\ref{EigEq}). The eigenvalue equations for the positive- and
negative-energy modes differ by the sign of the energy. Alternatively, we
can take the negative-energy modes in the form (\ref{psim}). With this
representation, the eigenvalue equation for the negative-energy modes is
obtained from the positive-energy one by inverting the sign of the parameter
$\alpha $, the latter being the ratio of the magnetic flux to the flux
quantum.

The mode-sums for VEVs of the charge and current densities, Eqs. (\ref{j0})
and (\ref{j2}), contain series over the roots of Eq. (\ref{EigEq}). The
latter are given implicitly and these representations are not well adapted
for the investigation of the VEVs. More convenient expressions are obtained
by making use of the generalized Abel-Plana formula (\ref{Sum}) for the
summation of the series. The formulas obtained in this way have two
important advantages: the explicit knowledge of the eigenvalues is not
required and the boundary-induced contributions to the VEVs are explicitly
extracted. In addition, instead of series with highly oscillatory terms for
large values of quantum numbers, in the new representation one has
exponentially convergent integrals for points away from the edges. This is
an important point from the point of view of numerical evaluations.

The VEVs for both the charge and current densities are decomposed into
boundary-free, single boundary-induced and the second boundary-induced
contributions. All them are odd periodic functions of the magnetic flux with
the period equal to the flux quantum. For the geometry with two boundaries
we have provided two representations, given by Eqs. (\ref{j02bn}) and (\ref%
{j02bnp}) for the charge and by Eqs. (\ref{j22n}) and (\ref{j2b}) for the
azimuthal current. In these representations the contributions for the
exterior or interior geometries with a single boundary are explicitly
extracted. The last terms in all the representations are induced by the
introduction of the second boundary to the geometry with a single boundary.
The single boundary parts in the VEVs are given by the expressions (\ref%
{j0a2}) and (\ref{j2abn}) in the exterior region and by Eqs. (\ref{j0b2n})
and (\ref{j22b}) for the interior region.

Unlike the case of the boundary-free geometry the charge and current
densities in the ring are continuous at half-odd integer values for the
ratio of the magnetic flux to the flux quantum, and both of them vanish at
these points. We have shown that the behaviour of the VEVs as functions of
the field mass (energy gap in field theoretical models of planar condensed
matter system) is essentially different for the cases $s=1$ and $s=-1$. With
the initial increase of the mass from the zero value, the modulus for the
charge and current densities decreases for the irreducible representation
with $s=1$ and increases for the one with $s=-1$. With further increase of
the mass the vacuum densities are suppressed in both cases. An important
feature that distinguishes the VEVs of the charge and current densities from
those for the energy-momentum tensor is their finiteness on the boundaries.
On the outer edge the current density is equal to the charge density whereas
on the inner edge they have opposite signs. For a fixed values of the other
parameters, both the charge and current densities decrease by the modulus
with decreasing outer radius.

The boundary condition (\ref{BCMIT}) we have considered contains no
additional parameters and is a special case in a general class of boundary
conditions for the Dirac equation confining the fermionic field in a finite
volume. It is the most popular boundary condition in the investigations of
the fermionic Casimir effect for various types of the bulk and boundary
geometries. On the base of the analysis given above we can consider another
boundary condition that differs from (\ref{BCMIT}) by the sign of the term
containing the normal to the boundary. The corresponding expressions for the
VEVs of the charge and current densities are obtained from those in sections %
\ref{sec:Charge} and \ref{sec:Current} by making the replacement (\ref%
{nurepl}) for $n_{u}$ defined by Eq. (\ref{nab}). All the final formulas
(for example, Eqs. (\ref{j02bn}), (\ref{j2b})) remain the same with the only
difference in the definition of the notation (\ref{fnun}), where now $n_{u}$
should be replaced by $-n_{u}$. Equivalently, the results for the field with
a given $s$ and with the modified boundary condition are obtained from the
corresponding expressions for the field with $-s$ and obeying the condition (%
\ref{BCMIT}) replacing $f_{n_{p}}^{(u)}(z)$ by its complex conjugate, $%
f_{n_{p}}^{(u)\ast }(z)$. With the modified boundary condition, the current
density is equal to the charge density on the inner edge and has the
opposite sign on the outer edge.

The charge and current densities in parity and time-reversal models are
obtained combining the results for the separate cases with $s=1$ and $s=-1$.
These models can be formulated in terms of four-component spinors
constructed from the 2-component spinors realizing the two different
irreducible representations. Assuming that both these spinors obey the
boundary condition (\ref{BCMIT}) and have the same mass, the resulting
charge density vanishes, whereas the current density is obtained from the
expressions given in section \ref{sec:Current} with the additional factor 2.
For the graphene circular rings, an additional factor 2 comes from the spin
degree of freedom.

\section*{Acknowledgments}

A.A.S. was supported by the State Committee of Science Ministry of Education
and Science RA, within the frame of Grant No. SCS 15T-1C110, and by the
Armenian National Science and Education Fund (ANSEF) Grant No. hepth-4172.
The work was partially supported by GRAPHENE-Graphene-Based Revolutions in
ICT and Beyond, project n.604391 FP7-ICT-2013-FET-F, as well as by the NATO
Science for Peace Program under grant SFP 984537.

\end{document}